\documentclass[prx,twocolumn,superscriptaddress,floatfix,nopacs]{revtex4}
\usepackage{graphicx,amsfonts,amssymb,amsmath,enumerate}

\usepackage{hyperref}
\usepackage{cleveref}

\usepackage{xcolor}

\newif\ifhyper
\hypertrue
\ifhyper
\hypersetup{
   citecolor = {green},
   colorlinks = {true}, 
   urlcolor = {blue} 
}
\fi

\newcommand{\beq}{\begin{equation}}
\newcommand{\eeq}{\end{equation}}
\newcommand{\beqa}{\begin{eqnarray}}
\newcommand{\eeqa}{\end{eqnarray}}
\newcommand{\ket} [1] {\vert #1 \rangle}
\newcommand{\bra} [1] {\langle #1 \vert}

\newcommand{\tr}{\mathop{\mathrm{tr}}}

\def\bra#1{\langle#1\vert}
\def\ket#1{\vert#1\rangle}

\def\Longarrow{\protect\@lra}
\def\@lra{\relbar\joinrel\relbar\joinrel\relbar\joinrel%
          \relbar\joinrel\rightarrow}

\DeclareMathOperator{\arcsinh}{arcsinh}

\begin{document}

\title{Holographic encoding of universality in corner spectra}

\author{Ching-Yu Huang}
\affiliation{C. N. Yang Institute for Theoretical Physics and Department of Physics and Astronomy, State University of New York at Stony Brook, NY 11794-3840, USA}

\author{Tzu-Chieh Wei}
\affiliation{C. N. Yang Institute for Theoretical Physics and Department of Physics and Astronomy, State University of New York at Stony Brook, NY 11794-3840, USA}

\author{Rom\'an Or\'us}
\affiliation{Institute of Physics, Johannes Gutenberg University, 55099 Mainz, Germany}

\begin{abstract}
In numerical simulations of classical and quantum lattice systems, 2d corner transfer matrices (CTMs) and 3d corner tensors (CTs) are a useful tool to compute approximate contractions of infinite-size tensor networks. In this paper we show how the numerical CTMs and CTs can be used, {\it additionally\/}, to extract universal information from their spectra. We provide examples of this for classical and quantum systems, in 1d, 2d and 3d. Our results provide, in particular, practical evidence for a wide variety of models of the correspondence between $d$-dimensional quantum and $(d+1)$-dimensional classical spin systems. We show also how corner properties can  be used to pinpoint quantum phase transitions, topological or not, without the need for observables. Moreover, for a chiral topological PEPS we show by examples that corner tensors can be used to extract the entanglement spectrum of half a system, with the expected symmetries of the $SU(2)_k$ Wess-Zumino-Witten model describing its gapless edge for $k=1,2$. We also review the theory behind the quantum-classical correspondence for spin systems, and provide a new numerical scheme for quantum state renormalization in 2d using CTs. Our results show that bulk information of a lattice system is encoded holographically in efficiently-computable properties of its corners.  

\end{abstract}

\maketitle

\section{Introduction}  \label{sec:intro}

Corner Transfer Matrices (CTMs) were introduced by Baxter in the context of exactly solvable models in 2d.  
{ In his 1968 paper \cite{baxterCTM} he laid, without noticing it, some of the basics of CTMs, together with those of  the density matrix renormalization group (DMRG) and matrix product states (MPS), when dealing with the dominant eigenvector of a 1d transfer matrix.} 
CTMs are a key ingredient in the exact solution of several statistical-mechanical models \cite{CTMstat}, and have also inspired many advances in the study of quantum many-body entanglement \cite{PeschelCTM, dirCTM, 3dCT}.
CTMs have also been important for the numerical simulation of lattice systems, both classical and quantum. 
{ In retrospect,  Baxter proposed in 1978 a variational method over CTMs \cite{BaxterVarCTM}, inspired by an earlier numerical method from 1941 by Kramers and  Wannier \cite{KWappro}.}
This, in turn, was one of the inspirations of Nishino and Okunishi's  CTM Renormalization Group method (CTMRG) \cite{CTMRG}. 
(For the avid reader, a good source of information about this history can be found in Ref.\cite{ctmnote}.)
Similar numerical CTM techniques are also currently used in the calculation of low-energy properties of infinite-size quantum lattice systems in 2d \cite{dirCTM, ffUpdate, frankCTM}, for which they have become one of the standard tools in the approximate calculation of physically-relevant quantities such as expectation values of local observables and low energy excitations. CTMs and their algorithms have also been generalized to 3d by the so-called Corner Tensors (CTs) \cite{CT, 3dCT}, in turn allowing to explore higher-dimensional systems with Tensor Network (TN) methods. 

Still, CTMs and CTs contain a great amount of holographic information about the bulk properties of the system which, a bit surprisingly, has not yet been fully exploited in the context of numerical simulations. Apart from being a useful object in the calculation of observables, the corner objects also contain, by themselves, information about the universal properties of the simulated model, providing a nice instance of the bulk-boundary correspondence for Tensor Networks (TNs) \cite{tn}. Bulk information is encoded holographically at the ``boundary" corners, in a way similar to the study of the so-called ``entanglement spectrum" and ``entanglement Hamiltonians" \cite{eHam}. 
For instance, Peschel, Kaulke and Legeza~\cite{PeschelCTM} showed that the entanglement spectrum of a quantum spin chain (w.r.t. a partition into two semi-infinite segments) is identical, up to some normalization constant, to the spectrum of some CTM in 2d, which could be computed exactly in some cases. This was the case of the Ising and Heisenberg quantum spin chains in a transverse field, for which they were able to compute such a entanglement spectrum exactly as eigenvalues of a ``corner Hamiltonian", which here we call ``corner energies". Nevertheless, and in spite of these results, the study of the physical information encoded holographically in CTMs and CTs has been traditionally overlooked in numerical simulations, especially in the case of 2d quantum lattice systems, in spite of the fact that this is, indeed, a quite natural thing to do. 

In this paper we explore the fingerprints of universal physics that are encoded holographically in numerical CTMs and CTs. We do this by studying the eigenvalue spectra of these objects or, more precisely, of \emph{contractions} of these objects, together with its associated entropy, in a way to be explained later. We provide several examples of this both for classical and quantum systems, including classical and quantum Ising, XY, XXZ and $N$-state Potts models, as well as several instances of 2d Projected Entangled Pair States (PEPS) \cite{PEPS} describing perturbed $\mathbb{Z}_2$, $\mathbb{Z}_3$, symmetry-protected, and chiral topological orders \cite{Z2, IsingPEPS, Z3_1, Z3_2, spt, chiralPEPS, chiral1, chiral2}. To achieve this goal we use a variety of TN methods for CTMs and CTs. For the case of ground-state properties of a quantum Hamiltonian $H_q$ in $d$ dimensions, we set up a corner method for a $d+1$ dimensional TN as described in Ref.~\cite{3dCT} via the imaginary-time evolution operator $e^{-\tau H_q}$ for large enough $\tau$. From a broad perspective, some of our results can be understood as a generalization of the work by Peschel, Kaulke and Legeza \cite{PeschelCTM} to 2d quantum systems. Additionally, whenever we have direct access to the ground-state wavefunction $|\psi_G\rangle$ in the form of a TN (e.g., a PEPS), we can also study the CTMs originating from the TN for the norm $\langle \psi_G|\psi_G\rangle$, which can be regarded as the partition function of some fictitious 2d classical model with complex weigths. Throughout this paper we shall refer to this setup as {\it reduction} CTM (rCTM), since it is a scheme that ``reduces" the wavefunction to a partition function. Such CTMs are, in fact, readily available in several TN algorithms (such as the full update and fast full update for infinite PEPS \cite{iPEPS, ffUpdate}). Along the way, we also compare different schemes for the classical-quantum correspondence, and provide some pedagogical derivations.

When the quantum state $|\psi_G\rangle$ is explicitly given by a TN, we can directly obtain its associated CTs. To do this we propose a new scheme for quantum state renormalization. In this case, the entanglement spectrum of a partition (of infinite size) can be readily obtained by diagonalizing a contraction of CTs, as we shall explain.  
{ First we use the Ising PEPS in the disorder phase as an example to demonstrate how to obtain the CTs entanglement spectrum. }
Then we also apply this quantum state renormalization to two cases of chiral topological ordered states, with  $SU(2)_k$ edge modes (for $k=1,2$), and find the degeneracy pattern in the entanglement spectrum matches that in the corresponding conformal tower for the vacuum of the $SU(2)_k$ WZW model. 

Our work is organized as follows: in Sec.(\ref{sec:CornerIntro}) we provide a reminder on CTMs, CTs,  some of their properties, as well as a summary of previous relevant results. In Sec.(\ref{Sec3}) we provide a summary of the TN numerical methods used to study the 1d, 2d and 3d classical and quantum lattice systems explored in this paper. Moreover, we also provide a new numerical scheme for quantum state renormalization in 2d using CTs. In Sec.(\ref{Sec4}) we analyze, as a first test, several models in the universality class of the quantum Ising spin chain in a transverse field. In Sec.(\ref{Sec5}) we show how the quantum-classical correspondence can be identified from corner properties, for 1d quantum vs 2d classical and 2d quantum vs 3d classical models. In this section we also review the theory behind several approaches for the quantum-classical correspondence, namely, the partition function approach, Peschel's approach, and Suzuki's approach for the XY model \cite{SuzukiXY}. Then, in Sec.(\ref{Sec6}) we provide further examples where the calculation of corner properties is useful. In particular, we show how corner properties can be used to pinpoint phase transitions in quantum systems ``almost for free" in common tensor network numerical algorithms, without the need to compute observables explicitly. We show this for several PEPS with topological order, including symmetry-protected, as well as for the 2d XXZ model. In Sec.(\ref{Sec7}) we show how CTs can be used to compute the entanglement spectrum of several bipartitions of an infinite 2d system. In particular, we apply the idea to chiral topological PEPS \cite{chiral1}, showing that the obtained spectra encode the expected symmetries of the chiral conformal field theory (CFT) describing its gapless edge, specifically, $SU(2)_k$ WZW models for $k=1,2$. Finally, in Sec.(\ref{sec:Conclusion}) we wrap up with a summary of the results, conclusions and perspectives. 

\section{Corner objects}  \label{sec:CornerIntro}

\subsection{Corner transfer matrices}
\begin{figure}
\includegraphics[width=0.5\textwidth]{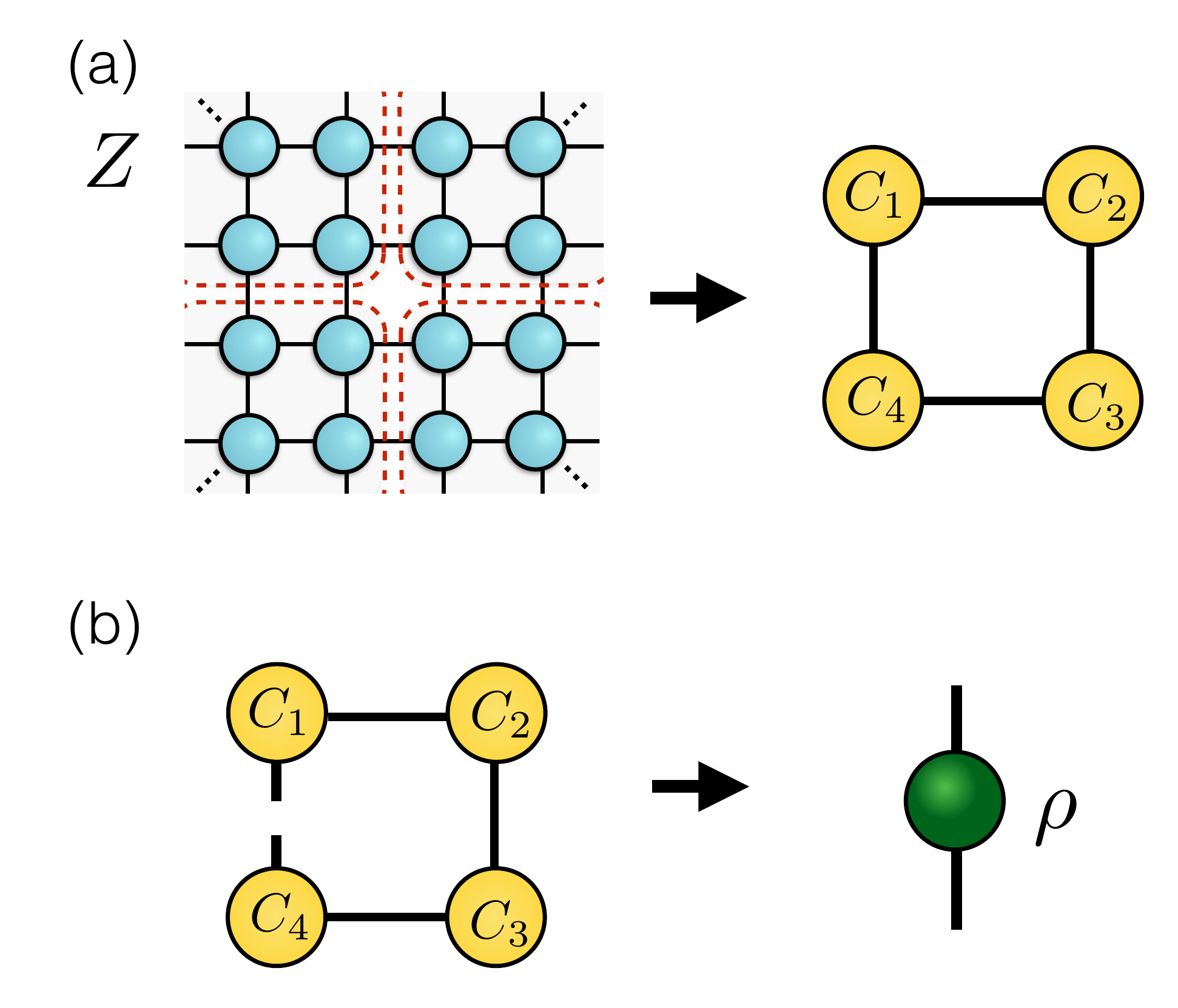}
\caption{ [Color online] (a) The contraction of a 2d square lattice of tensors results in a scalar $Z$, understood as the trace of the product of four CTMs, one for each corner. (b) A reduced density matrix $\rho$ of a system with a CTM at every corner.}  
\label{fig:CTMmethod}
\end{figure}

CTMs are objects that can be defined for any 2d tensor network. Here, for simplicity, we assume the case of a 2d TN on a square lattice. Such a TN could be, e.g., the partition function of a classical lattice model, the time-evolution of a 1d quantum system, or the norm of a 2d PEPS. To define what a CTM is, we notice that the contraction of the 2d TN can be obtained, at least theoretically, by multiplying four matrices $C_1,C_2,C_3$ and $C_4$, one for each corner (see Fig.~\ref{fig:CTMmethod}a). Therefore, one has that  
\begin{equation}
Z = {\rm tr} \left(C_1C_2C_3C_4 \right) , 
\label{ctmm}
\end{equation}
where $Z$ is the scalar resulting from the contraction. Matrices $C_1,C_2,C_3$ and $C_4$ are the \emph{Corner Transfer Matrices} of the system. They correspond to the (sometimes approximate) contraction of all the tensors in each one of the four corners of the 2d TN. In some cases, when the appropriate lattice symmetries are present, the four CTMs are equal, i.e., $C \equiv C_1 = C_2 = C_3 = C_4$. For the sake of simplicity, in this section  we shall assume that this is the case, though in the following sections the four CTMs are different when computed numerically. 

It is also convenient to define diagonal CTMs $C_d = P C P^{-1}$. Depending on the symmetries of the system (and thus of $C$), matrix $P$ may be arbitrary, unitary or orthogonal. Let us call the eigenvalues $\nu_{\alpha}$, with $\alpha = 1, 2, \ldots, \chi$, and $\chi$ the \emph{bond dimension} of the CTM. Then, the contraction of the full TN reads 
\begin{equation}
Z = {\rm tr} \left(C_d^4 \right) = \sum_{\alpha=1}^\chi \nu_{\alpha}^4 .
\label{ctm2}
\end{equation}
In fact, one can understand this as the trace of the exponential of a ``corner Hamiltonian" $H_C$, i.e.,
\begin{equation}
Z = {\rm tr} \left(e^{-H_C} \right),   
\label{ctm3}
\end{equation}
with  
\begin{equation}
H_C \equiv - \log{ \left(C_d^4 \right) }.    
\label{ctm4}
\end{equation}
Notice that a similar Hamiltonian can also be defined individually for each one of the corners.  

Depending on the symmetries of the CTMs, $H_C$ may be a Hermitian operator or not.  From the point of view of quantum states of 1d quantum lattice systems, it is well known \cite{3dCT} that operator $e^{-H_C}$ is related to the reduced density matrix of half an infinite chain (with $H_C$ Hermitian in this case), see Fig.~\ref{fig:CTMmethod}b. In fact, the spectrum of Schmidt coefficients $\lambda_\alpha$ of half an infinite quantum chain in its ground state is given by $\lambda_{\alpha} = \nu_{\alpha}^2$. These Schmidt coefficients are related to the eigenvalues $\omega_\alpha$ of the reduced density matrix of half an infinite quantum system (the so-called ``entanglement spectrum" \cite{eHam}) by $\omega_\alpha = \lambda_\alpha^2 = \nu_\alpha^4$, which are known to codify universal information about the system when close enough to criticality \cite{PeschelCTM}. In terms of $\omega_\alpha$, the contraction of the 2d TN reads $Z = \sum_{\alpha = 1}^\chi \omega_\alpha$. Aditionally, the eigenvalues $\varepsilon_\alpha$ of the corner Hamiltonian $H_C$ read 
\beq
\varepsilon_\alpha \equiv - \log \omega_\alpha.
\eeq
In this paper we call these eigenvalues $\varepsilon_\alpha$'s \emph{corner energies}.

\subsection{Corner Tensors}
\begin{figure}
\centering
\includegraphics[width=0.45\textwidth]{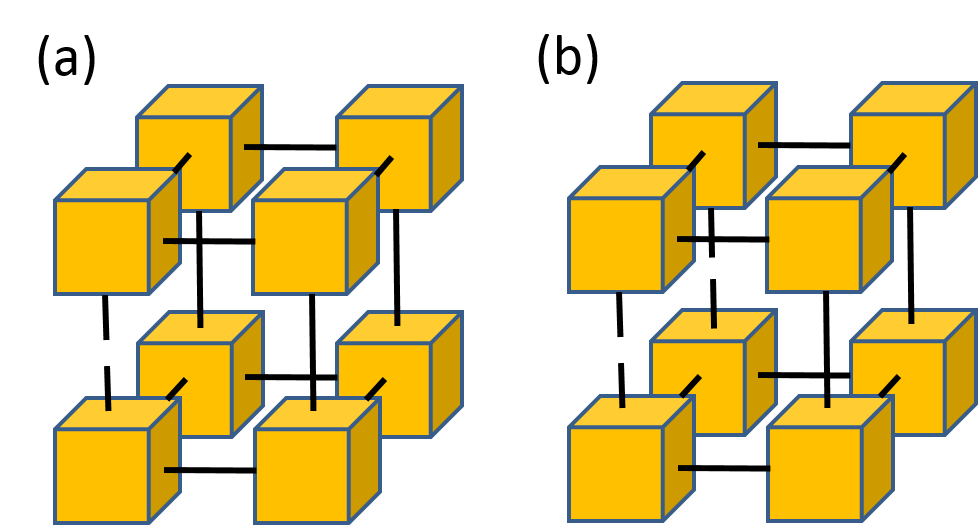} 
\caption{[Color online] 3d corner tensors which correspond to tracing over, respectively, (a) three quarters and (b) half of a given 2d quantum system.} 
\label{fig:CT}
\end{figure}

Similarly to CTMs for 2d TNs, one can define corner objects for higher dimensions, which we generically call \emph{Corner Tensors} (CT). Formally speaking, a CT is the (sometimes approximate) contraction of all the tensors at one of the corners of a TN. For instance, for a TN on a 3d cubic lattice, one would have that its contraction $Z$ is equivalent to the contraction of eight CTs, i.e.,  
\beq
Z = f(C_1, C_2, C_3, C_4, C_5, C_6, C_7, C_8), 
\eeq
with $C_i$ ($i=1, \ldots, 8$) eight three-index tensors (the CTs), and $f(\cdot)$ a function specifying the contraction pattern, see Fig.~\ref{fig:CT}. 

For the case of systems with CTs it is also possible to define corner Hamiltonians. For instance, contractions such as the ones in Fig.~\ref{fig:CT} correspond, for the case of a 2d quantum lattice system, to tracing over three quarters or half of the infinite system. For quantum systems described by a 2d PEPS, it is possible to obtain these types of contractions by using the quantum state renormalization scheme from Sec.(\ref{Sec3}). In such cases, these contractions correspond to the reduced density matrices $\rho$ of either one quarter or half an infinite 2d system, with eigenvalues $\omega_\alpha$, $\alpha = 1, \ldots, \chi$ (entanglement spectrum). The contraction of the full 3d TN thus amounts to $Z = \sum_{\alpha = 1}^\chi  \omega_\alpha$, as in the lower-dimensional case of CTMs. Again, it is possible to define a corner Hamiltonian $H_C$ and corner energies $\varepsilon_\alpha$ in an analogous way as for CTMs. 

\subsection{Previous results}

CTMs and CTs have proven to be important in a variety of contexts, both for theory and numerics. In  statistical mechanics they were used to solve the hard hexagon model and many others \cite{baxterCTM, CTMstat}. From the perspective of quantum information, it is well known that the  corner Hamiltonian $H_C$ is related to a quantum system which, in some cases, can be diagonalized exactly \cite{PeschelCTM}. Numerically, Baxter developed a variational method to approximate the partition function per site of a 2d classical lattice model by truncating in the eigenvalue spectrum of the CTM \cite{BaxterVarCTM}. This was later refined by Nishino and Okunishi, who developed the Corner Transfer Matrix Renormalization Group method (CTMRG) \cite{CTMRG}.  Alternative truncation schemes for CTMRG have also been studied, based on a directional approach and with a direct application in infinite-PEPS algorithms \cite{iPEPS, dirCTM}. In fact, CTMs have been applied extensively in the calculation of effective environments in infinite-PEPS simulations \cite{oruscorboz}. Moreover, they have been used as well in the generalization to 2d of the time-dependent variational principle \cite{frankCTM}, which is also useful in the calculation of 2d excitations. As for generalizations, CTMs have also been used in other 2d geometries, including lattice discretizations of AdS manifolds \cite{ads}. Numerical methods with CTMs were also implemented in systems with periodic boundary conditions \cite{pbc} as well as  stochastic models \cite{stoch}. Methods targeting directly the corner Hamiltonian have also been considered \cite{Hc, Kim2016}. Finally, the higher-dimensional generalization to corner tensors has also been used to develop new numerical simulation algorithms \cite{CT, 3dCT}. 

\section{Approach and methods}
\label{Sec3}

\subsection{Generalities}

In the following sections we shall show how the spectrum of eigenvalues $\omega_\alpha$, or equivalently the spectrum of corner energies $\varepsilon_\alpha$, encodes useful universal information when computed numerically for a variety of classical and quantum lattice systems. This is also true for the ``corner entropy" given by 
\beq
S \equiv - \sum_\alpha \omega_\alpha \log \omega_\alpha . 
\eeq
In particular, we will show explicitly how the spectrum as well as the entropy exactly coincide if compared between some $d$-dimensional quantum and $(d+1)$-dimensional classical spin systems, as expected from the quantum-classical correspondence. Moreover, we will also study  them for a variety of other models, including several instances of topologically-ordered states. We will see that this can be useful to pinpoint phase transitions as well as to study edge physics of chiral topological states. 

Concerning numerical algorithms, in our simulations we have used the following, depending on the nature of the system to be studied: 

\begin{figure}
\includegraphics[width=0.5\textwidth]{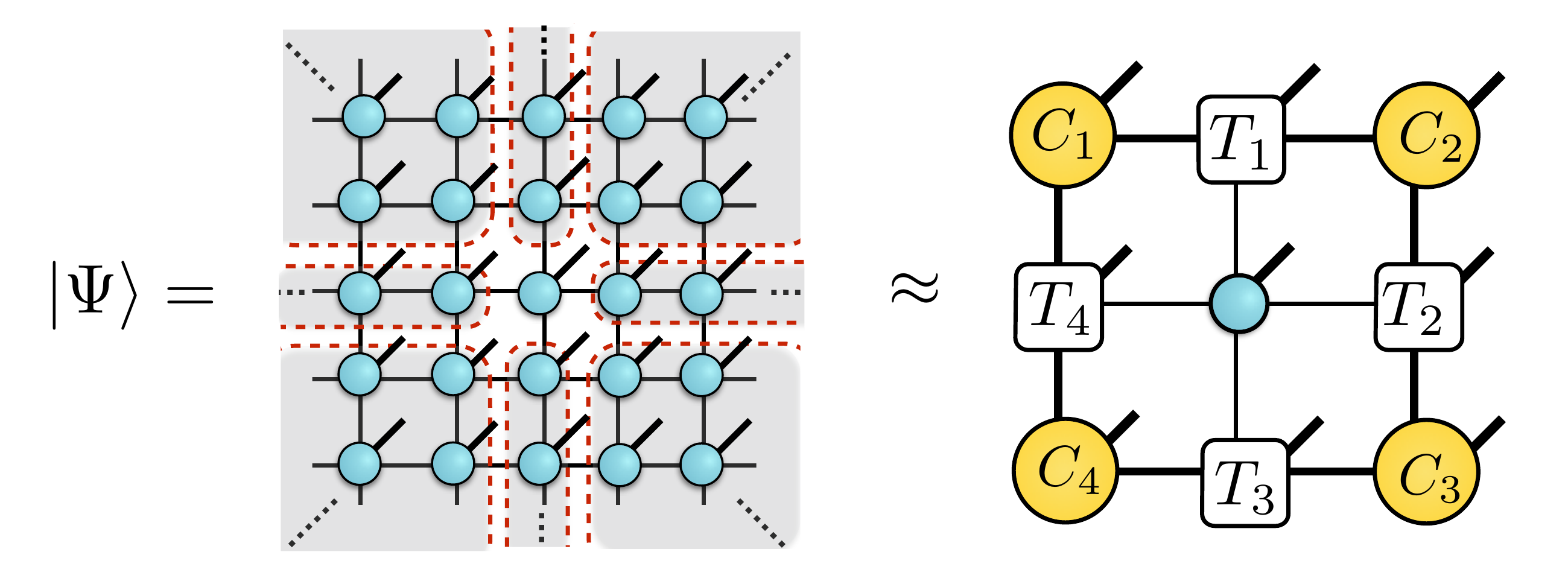}
\caption{[Color online] 2d PEPS on a square lattice and its renormalized version with CTs}
  \label{fig:effective_WF}
\end{figure}
\begin{figure}
\includegraphics[width=0.5\textwidth]{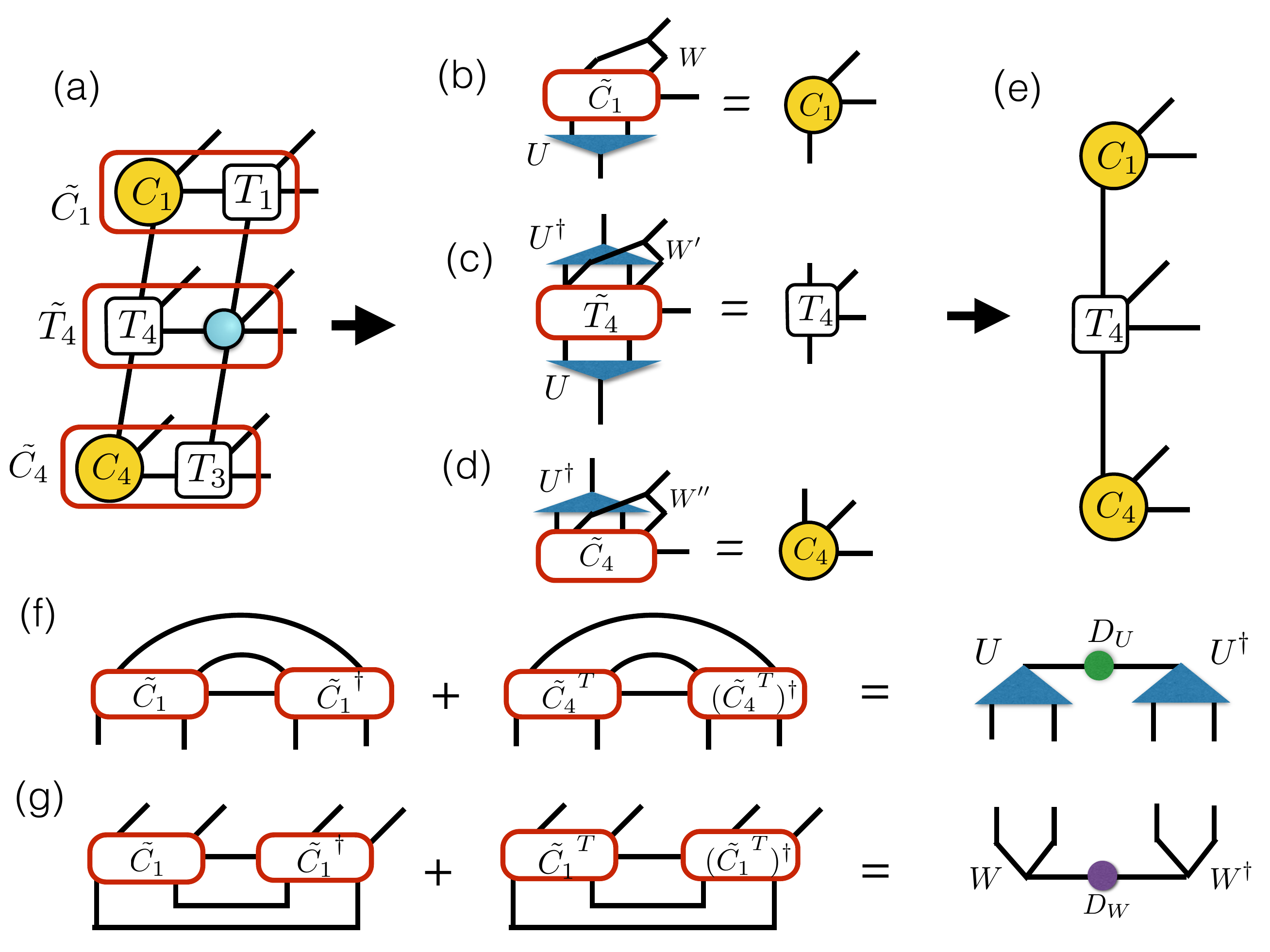}
\caption{[Color online] 2d quantum state renormalization with corner tensors: a left move, where one column is absorbed to the left. The procedure is the same as in the directional CT approach from Ref.~\cite{3dCT}, but on a single layer of PEPS tensors instead of two layers. Consequently, at every step we need to renormalize with isometries not just the bond indices, but also the physical indices, which proliferate at every iteration. 
Several prescriptions are possible for the calculation of the isometries, e.g, one could consider higher-order  singular value decompositions of the resulting tensors \cite{genSVD}, or compute the reduced density operators of the indices to be truncated \cite{CTMRG}.} 
\label{fig:qsrg_ctm}
\end{figure}

\begin{enumerate} 
\item{\emph{For 1d quantum:} the infinite Time-Evolving Block Decimation (iTEBD) \cite{itebd} to approximate ground states. The spectrum $\omega_\alpha$ obtained from CTMs is easily related \cite{PeschelCTM} to the Schmidt coefficients $\lambda_\alpha$ of a bipartition, readily available from iTEBD or iDMRG \cite{idmrg}, as $\omega_\alpha = \lambda_\alpha^2$. In some instances we also use the simplified one-directional 1d method from Ref.~\cite{3dCT}.}
\item{\emph{For 2d classical:} 2d directional CTM approach \cite{dirCTM}.  }
\item{\emph{For 2d quantum:}  if a quantum Hamiltonian is given, then we use the 3d directional CTM approach~\cite{dirCTM} to compute properties of CTs, as well as infinite-PEPS (iPEPS) \cite{iPEPS} to approximate ground states. If the ground state $\ket{\psi_G}$ is given, then we use the directional CTM approach for the double-layer tensors of the norm \cite{dirCTM} to compute 
the ``reduced" spectrum $\omega^{(r)}_\alpha$ from rCTM. Moreover, we also use the 2d quantum state renormalization described in the next section to compute  properties of CTs. As we shall see, this method is single-layer and targets directly the quantum state.}
\item{\emph{For 3d classical:} simplified one-directional 2d method~\cite{3dCT}.}
\end{enumerate} 

\subsection{2d quantum state renormalization with CTs} 
\label{qsr}

The procedure of quantum state renormalization is important in 2d to obtain the contractions from Fig.~\ref{fig:CT} in the quantum case, which give the reduced density matrix by tracing spins in three quadrants or half-infinite plane. The entanglement spectrum can then be obtained from the eigenvalues of such reduced density matrix. We have implemented our own approach for the case of a 2d PEPS, using CTs and single-layer contractions. This procedure, which is an independent algorithm by itself, is explained in detail in what follows. 

The quantum state renormalization group (QSRG) transformation acts directly on a quantum state and aims to extract a fixed-point wave function encoding universal properties \cite{qsrg}. The basic idea is to remove non-universal short-range entanglement related to the microscopic details of the system. After many rounds of QSRG, the original ground state flows to a simpler fixed-point state, from which one can identify to which phase the system belongs to.

In order to determine the fixed-point wave function we make use of CTs, see Fig.~\ref{fig:effective_WF}. The distinction from the usual QSRG is that here the fixed-point wave function will be encoded in these CTs.
The procedure is similar to the directional CTM approach from Ref.~\cite{dirCTM}, but this time acting directly on the PEPS, which is single-layer, and not on the TN for the norm, which is double-layer. An example of a left-move is in Fig.~\ref{fig:qsrg_ctm}, where we show also a simple option to obtain the isometrics needed for the coarse-grainings. We follow this procedure by absorbing rows and columns towards the left, up, right and down directions until convergence is reached. In the end, the corner tensors $C$ represent the renormalization of one quadrant of the 2d PEPS, and the half-row/half-column tensors $T$ to the renormalization of half an infinite row or column of tensors in the PEPS. One then follows the contractions in Fig.~\ref{fig:CT} to obtain the corresponding reduced density matrix and hence the entanglement spectrum.

\section{First test: the 1d quantum Ising universality class}
\label{Sec4}

In order to build some intuition about the numerical information contained in the spectrum $\varepsilon_\alpha$ of corner energies, we have first performed a series of numerical tests in systems belonging to the universality class of the 1d quantum Ising model in a transverse field. The analyzed models undergo a 2nd order quantum or classical phase transition, with the critical point being described by an effective $(1+1)$-dimensional CFT of a free fermion \cite{cftFreeFermion}. The models and methods considered are: 


\medskip\underline{\emph{(i) 1d quantum Ising:}} the quantum Hamiltonian is given by  
\beq
H_q= - \sum_i \sigma_x^{[i]} \sigma_x^{[i+1]} - h \sum_i \sigma_z^{[i]}, 
\eeq
with $\sigma_\alpha^{[i]}$ the corresponding $\alpha$-Pauli matrices at site $i$, and $h$ the transverse magnetic field, with critical point at $h_c = 1$. We used iTEBD to approximate the ground state by a Matrix Product State (MPS) \cite{mps} and here the square of the Schmidt coefficients  $\lambda^2_\alpha$ (hence the entanglement spectrum) is obtained.  We also 
use the simplified one-directional 1d method from Ref.~\cite{3dCT} to obtain the corner spectrum $\omega_\alpha$. As argued in Ref.~\cite{PeschelCTM} we expect and verify that $\{\lambda_\alpha^2\}$ agrees with $\{\omega_\alpha\}$. 


\medskip\underline{\emph{(ii) 2d classical Ising:}} the partition function is given by 
\beq
Z_c = \sum_{\{ s \}}e^{- \beta H_c(\{ s \})}, 
\eeq
with classical Hamiltonian 
\beq
H_c\{s \}= -\sum_{\langle i, j \rangle } s^{[i]} s^{[j]}, 
\label{cIsing} 
\eeq
where $\beta = 1/T$ is the inverse temperature, $s^{[i]} = \pm 1$ is a classical spin variable at site $i$, $\{ s \}$ is a spin configuration, and the sum in the Hamiltonian runs over nearest neighbours on the square lattice. The model is exactly solvable, and the critical point satisfies $\beta_c  = \frac{1}{2}\log { \left( 1+\sqrt{2} \right) }$. It is well known \cite{itebd} that the partition function $Z_c$ can be written as an exact 2d tensor network with tensors on the sites of a square lattice. The approximate contraction is therefore amenable to tensor network methods. We use the directional CTM approach to compute the corner spectra and corner entropy from the tensors defining the partition function of the model.


\medskip\underline{\emph{(iii) 2d Ising PEPS:}} as explained in Ref.~\cite{IsingPEPS}, it is actually possible to write an exact Projected Entangled Pair State (PEPS) \cite{PEPS} with bond dimension $D=2$ whose expectation values are the ones of the 2d classical Ising model. The way to construct this PEPS is simple: one starts by considering the quantum state 
\beq
\ket{\psi (\beta)}=\frac{1}{Z_c}e^{\left( \frac{\beta}{2} \sum_{\langle i,j \rangle} \sigma_z^{[i]} \sigma_z^{[j]} \right)} \ket{+, +, \cdots, +},  
\eeq
with $\beta$ some inverse temperature and $\ket{+}$ the $+1$ eigenstate of $\sigma_x$. It is easy to see that the expectation values of this quantum state match the ones of the 2d classical Ising model, e.g., 
\beq
\bra{\psi (\beta)} \sigma_z^{[i]} \sigma_z^{[j]} \ket{\psi (\beta)} = \frac{1}{Z_c} \sum_{ \{ s \} }s^{[i]} s^{[j]}  e^{-\beta H_c(\{ s \})} = \langle s^{[i]} s^{[j]} \rangle_\beta, 
\eeq
with $H_c(\{ s \} )$ the classical Hamiltonian in Eq.(\ref{cIsing}), and $\langle \cdot \rangle_\beta$ the expectation value in the canonical ensemble at inverse temperature $\beta$. 
For a square lattice, one can also see \cite{IsingPEPS} that the state $\ket{\psi (\beta) }$ can be written exactly as a 2d PEPS with bond dimension $D=2$. If $A$ is the tensor defining the PEPS, its non-zero coefficients are given by 
\beqa
A_{0000}^+ &=& \left( \cosh (\beta/2) \right)^4 \nonumber \\
A_{0010}^- &=& \left( \cosh (\beta/2) \right)^3  \left( \sinh (\beta/2) \right) \nonumber \\
A_{0110}^+ &=& \left( \cosh (\beta/2) \right)^2  \left( \sinh (\beta/2) \right)^2 \nonumber \\ 
A_{1110}^- &=& \left( \cosh (\beta/2) \right)  \left( \sinh (\beta/2) \right)^3 \nonumber \\
A_{1111}^+ &=& \left( \sinh (\beta/2) \right)^4  
\eeqa
and permutations thereof. In the above equations, the convention for the PEPS indices is $A_{\alpha \beta \gamma \delta}^i$, with $\alpha, \beta, \gamma, \delta$ the left, up, right and down indices, and $i$ the physical index (this time in the $+/-$ basis). By construction, this PEPS is critical at the same critical $\beta_c$ than the classical Ising model, and belongs also to the same universality class. For the numerical simulations it is sometimes convenient to parametrize the PEPS in terms of $g =  \frac{1}{2} \arcsin (e^{-\beta})$, and therefore $g_c \approx 0.349596$.  For this state, we computed the corner spectra and entropy from the double-layer TN defining its norm, using the directional CTM approach \cite{dirCTM}. 

\begin{figure}
\includegraphics[width=0.5\textwidth]{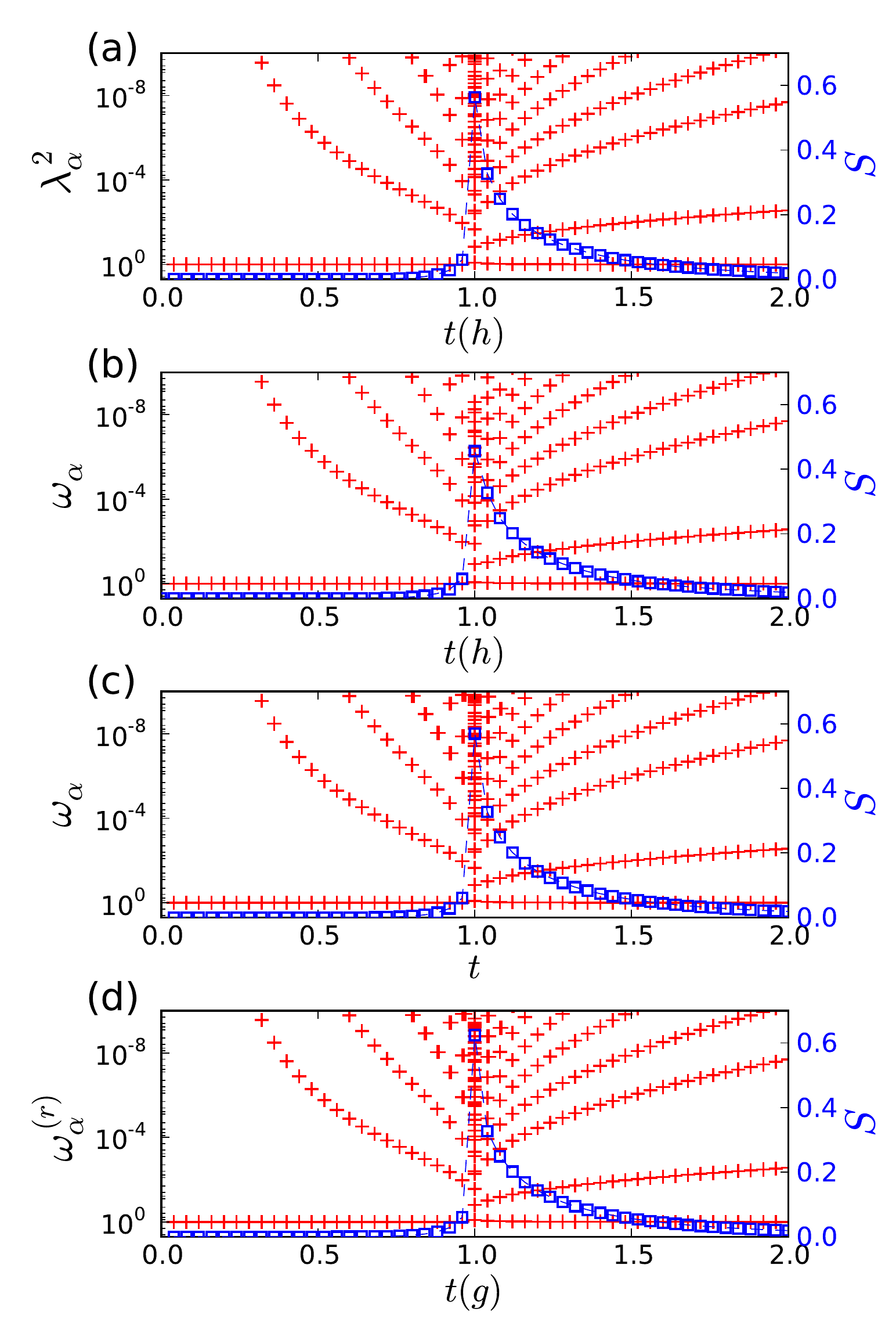}
\caption{[Color online] (a) Entanglement spectra $\lambda_{\alpha}^2$ and the entanglement entropy obtained from iTEBD of 1d quantum Ising model with parameter $t (h)$ as the function of  transverse field $h$.
The corner spectra $\omega_\alpha$ and the corner entropy $S$ of: (b) also the 1d quantum Ising model with parameter $t (h)$ as the function of  transverse field $h$, but computed with the simplified one-directional 1d method \cite{3dCT}; (c) 2d classical Ising model with temperature $t= T/T_c$; (d) 2d quantum Ising PEPS with parameter $t(g)$ as the function of $g$. In (c,d) the corner tensors are obtained from the rCTM setting, see also examples in Sec.~\ref{Sec6}. In all cases, the bond dimension of the CTMs - equivalent to the bond dimension of the MPS in case (a) - is $\chi=40$.}  
\label{fig:1dqising_2d}
\end{figure}

For these three models and the methods mentioned we have computed the spectrum $\omega_\alpha$ as a function of the relevant parameter (magnetic field, inverse temperature, perturbation...), as well as the corner entropy $S = - \sum_\alpha \omega_\alpha \log \omega_\alpha$.  
The results are shown in Fig.~\ref{fig:1dqising_2d}. 
The differences between models correspond to rescalings in the defining variables and parameters that map the different models among them. More specifically, we can rescale the parameters $h$ and $g$  using the 2d classical Ising reduced temperature $t=T/T_c$ as the basic variable, which is related to the magnetic field $h$ of the 1d quantum model by $t = T_c/\arcsinh  \sqrt{1/h}$, and to the parameter $g$ of the 2d Ising PEPS by $ t = -  T_c/ \log \left(\sin(2g) \right)$. As shown in the plots, in all cases one can see that the entropy $S$ tends to have the same type of divergence. Concerning the corner spectra $\omega_\alpha$, we see that all the models reproduce the same type of branches on both the symmetric and the symmetry-broken phases. 
As expected, all spectra match perfectly between the different calculations, since the different models can be mapped into each other exactly.

\section{Benchmarking the quantum-classical correspondence} 
\label{Sec5} 

In this section we  consider the corner energies for a variety of quantum and classical systems, which allows us to study in good detail the correspondence between quantum spin systems in $d$ dimensions and classical systems in $d+1$ dimensions. There are several approches and here we focus mainly on three of them, which we shall refer to as the partition-function method~\cite{partition}, Peschel's method~\cite{PeschelMap}, and Suzuki's method~\cite{SuzukiXY}, respectively.   We will give pedagoical treatment, specializing to a few models and show numerical results for a variety of 1d and 2d  quantum and 2d and 3d classical models.

\subsection{Partition-function approach}

We now review the standard procedure behind the partition-function approach for quantum-classical mapping and then examine such correspondence in terms of entanglement and corner spectra. The main idea is that, for a d-dimensional quantum Hamiltonian $H_q$ at inverse temperature $\beta$, the canonical quantum partition function $Z_q = \tr (e^{-\beta H_q})$  can be evaluated by writing it as a path integral in imaginary time, i.e., 
\begin{align}
Z_q &= \tr \left(e^{-\beta H_q} \right) = \sum_{m}  \langle m | e^{-\beta H_q}  | m \rangle, 
\end{align}
with $\ket{m}$ a given basis of the Hilbert space. Introducing resolutions of the identity at intermediate steps in imaginary time one has
\begin{align}
Z_q = \sum_{\{ m \} } \!  \langle m_0 | U | m_{L\!-1\!} \rangle  
\langle m_{L\!-1\!} | U | m_{L\!-2\!} \rangle \cdots  \langle m_{1} |U  | m_{0} \rangle, 
\end{align}
with $U \equiv e^{-\delta \tau H_q}$, $\delta \tau  \equiv \beta/L \ll 1$ (smaller than all time scales of $H_q$), and where the sum is for all the configurations of $m_\alpha, \alpha = 0, 1, \ldots, L-1$ with $m_L = m_0$, i.e., periodic boundary condition in imaginary time. 
 
As such, this way of writing the partition function can be interpreted in some cases as the one of a \emph{classical} model with some variables $m_\alpha$ along an extra dimension emerging from the imaginary-time evolution. In what follows, we make this specific for the quantum Ising and Potts models, and benchmark the theory with numerical simulations using CTMs and CTs computing the corner spectra and corner entropy. 

\subsubsection{Transverse field quantum Ising model in d dimensions} 

\emph{\underline{(i) Mapping via the partition function:}} let us consider the quantum Ising model with a transverse field in d dimensions for $L$ spins. For convenience, we use now the following notation for its Hamiltonian:
\begin{align}
H_q = -J_z \sum_{\langle i, j \rangle} \sigma_z^{[i]}  \sigma_z^{[j]} - J_x \sum_{i} \sigma_x^{[i]} 
=  H_z + H_x , 
\end{align}
where $\sigma_\alpha^{[i]} $ is the $\alpha$th Pauli matrix on site $i$, $J_z$ is the interaction coupling, $J_x$ the field strength, and the sum $\langle i, j \rangle$ runs over nearest-neighbors. The canonical quantum partition function of this model is given by
\begin{align}
Z_q &= \tr \left(e^{-\beta  H_q} \right) = \sum_{\eta_z }  \Big{\langle} \{\eta_z \}  \Big{|} e^{-\beta H_q }   \Big{|}  \{\eta_z \}  \Big{\rangle}, 
\end{align}
with $ \Big{|} \{ \eta_z \} \Big{\rangle}  \equiv   |  \eta_z^{[1]},  \eta_z^{[2]}, \cdots,    \eta_z^{[L]} \rangle $ the diagonal z-basis of the $N$ spins, so that $\eta_z^{[i]} = \pm 1, i = 1,2,...,L$. Splitting the imaginary time $\beta$ into infinitesimal time steps $\delta \tau$ we obtain 
\begin{align}
 & \Big{\langle}  \{ \eta_z(\tau +\delta \tau) \}   \Big{|}  e^{-\delta \tau H_q }  \Big{|} \{ \eta_z(\tau) \}  \Big{\rangle} \notag \\
 &  \approx 
 \Big{\langle} \{ \eta^z(\tau+\delta \tau) \}  \Big{|} e^{-\delta \tau H_x }    e^{-\delta \tau H_z }  \Big{|} \{ \eta^z(\tau) \}  \Big{\rangle} \notag \\
 &  =e^ { - \delta \tau  H_z (\{\eta_z(\tau) \} ) } 
  \Big{\langle} \{\eta^z (\tau+\delta \tau)\} \Big{|} e^{-\delta \tau H_x }  \Big{|} \{\eta^z(\tau)\} \Big{\rangle},  
  \label{split}
\end{align}
where in the first line we performed a first-order Trotter approximation with $O(\delta \tau^2)$ error.  Next, we consider the term with Hamiltonian $H_x$. In the single-site z-basis this can be written as  
\begin{align}
 & \langle \eta_z^{[i]} ({\tau+\delta \tau}) | e^{ \delta \tau   J_x  \sigma_x^{[i]} }  | \eta_z^{[i]}{(\tau)} \rangle \notag\\
 & = \sum_{\eta_x^{[i]}=\pm 1}  \langle \eta_z^{[i]} ({\tau+\delta \tau}) | e^{ \delta \tau  J_x  \sigma_x^{[i]}} |\eta_x^{[i]} \rangle \langle \eta_x^{[i]} | \eta_z^{[i]}{(\tau)} \rangle \notag\\
 &  = \sum_{\eta_x^{[i]}=\pm 1}  e^{  \delta \tau J_x \eta_x^{[i]} }   \langle \eta_z^{[i]} ({\tau+\delta \tau}) |\eta_x^{[i]} \rangle \langle \eta_x^{[i]} | \eta_z^{[i]}{(\tau)} \rangle. 
\end{align}
We can now use the overlap relation 
\beq
\langle \eta_x^{[i]} | \eta_z^{[i]} \rangle  = \frac{1}{\sqrt{2}}e^{ i \pi\left(\frac{1-\eta_x^{[i]}}{2} \right) \left( \frac{1-\eta_z^{[i]}}{2}  \right)}, 
\eeq
and define $\eta_z^{\prime [i]}  \equiv  \eta_z^{[i]} ({\tau+\delta \tau}) $,  $\eta_z^{[i]}  \equiv  \eta_z^{[i]} ({\tau}) $. Using this notation, we now have 
\begin{align}
 & \langle \eta_z^{\prime [i]} | e^{ \delta \tau J_x  \sigma_x^{[i]}}  | \eta_z^{[i]} \rangle \notag\\
  & = \sum_{\eta_x^{[i]}=\pm 1}   e^{ \delta \tau J_x \eta_x^{[i]} }  \times \frac{1}{2} 
 e^{ i \pi  \left( \frac{1-\eta_x^{[i]}}{2} \right) \left(\frac{1-\eta_z^{\prime [i]}}{2}  +   \frac{1-\eta_z^{[i]}}{2} \right)     } \notag \\
 & =  \frac{1}{2} \left( e^{ \delta \tau J_x}    + e^{ -\delta \tau J_x} \eta_z^{\prime [i]} \eta_z^{[i]}              \right) \notag \\
 & =  \frac{1}{2}    e^{ \delta \tau J_x}  \left( 1  + e^{ -2 \delta \tau J_x} \eta_z^{\prime [i]} \eta_z^{[i]}              \right).
 \label{CIrepresentation}
\end{align}
Moreover, we have the alternative representation 
\begin{align}
  \langle \eta_z^{\prime [i]} | e^{ \delta \tau J_x  \sigma_x^{[i]}   }  | \eta_z^{[i]} \rangle  &= C  e^{J_{\tau} \eta_z^{\prime [i]} \eta_z^{[i]}   } \notag \\
 & = C \left( \cosh (J_{\tau}) +  \sinh (J_{\tau})   \eta_z^{\prime [i]} \eta_z^{[i]}   \right)  \notag \\
 & = C  \cosh (J_{\tau}) \left(1 +   \tanh (J_{\tau})   \eta_z^{\prime [i]} \eta_z^{[i]}  \right), 
 \label{QIrepresentation}
\end{align}
with $C$ a normalization constant. Comparing Eqs.~(\ref{CIrepresentation}) and ~(\ref{QIrepresentation}), we obtain the relation 
$  \tanh (J_{\tau}) =   e^{ -2 \delta \tau J_x}$.  Finally, the partition function $Z_q$ of the transverse-field quantum Ising model can be written as 
\begin{align}
\label{eq:dQpartition}
Z_q  \approx  \sum_{ \{ \eta \}}  C'e^{J_s \sum_{\alpha, \langle i,j \rangle }  \eta_z^{[i]}(\tau_\alpha)  \eta_z^{[j]}(\tau_\alpha)} \nonumber \\
\times e^{J_{\tau}  \sum_{\alpha, i }  \eta_z^{[i]}(\tau_{\alpha + 1})  \eta_z^{[i]}(\tau_\alpha)}, 
\end{align}
where the ``coupling constants" along the imaginary-time ($\tau$) and space ($s$) directions are given by 
\beqa
J_{\tau} &=& \tanh^{-1}\left( e^{-2 \delta \tau J_x} \right) \nonumber \\
J_s &=& J_z \delta \tau . 
\eeqa

\begin{figure}
\includegraphics[width=0.3\textwidth]{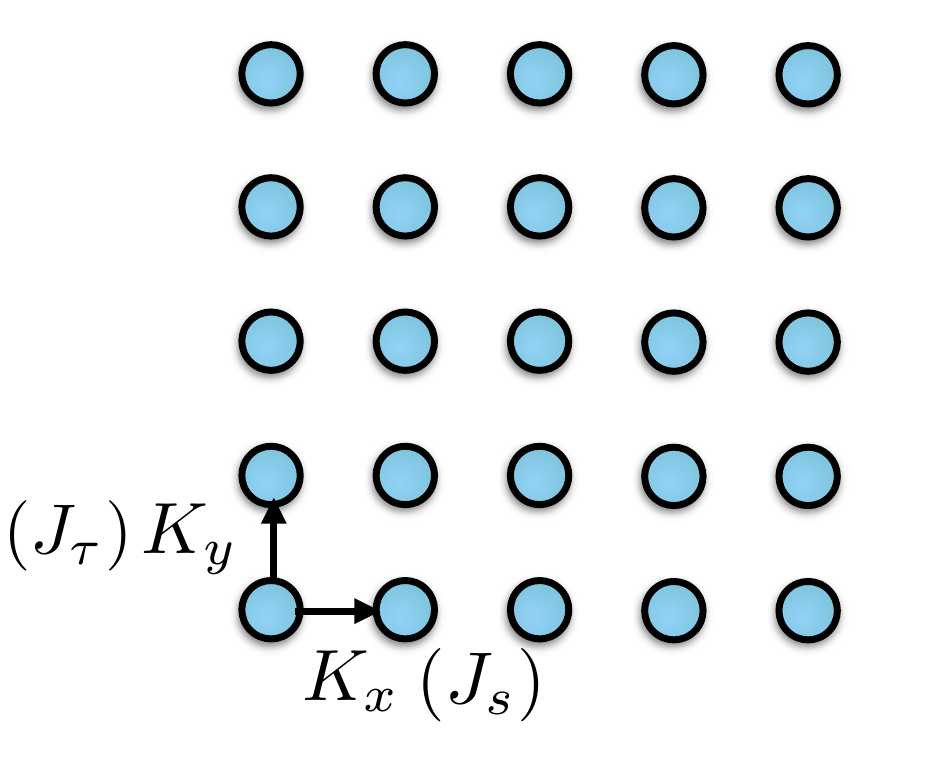}
\caption{[Color online] Coupling constants for a 2d classical Ising model. In connection with the quantum-classical correspondence, the vertical direction corresponds to imaginary-time.}  
\label{fig:2dlattice}
\end{figure}

Therefore, the canonical quantum partition function of a $d$-dimensional quantum Ising model with a transverse field at inverse temperature $\beta$ can be approximately represented by the classical partition function of a $(d+1)$-dimensional classical Ising model of size $\beta$ in the imaginary-time direction. 
The exact correspondence arrives if we take the number of sites $L$ in the imaginary time drection to be infinity, giving  $\delta=\beta/L \to 0$, and then the corresponding classical model has the couplings  $J_s\to 0$ and $J_\tau\to \infty$. In Monte Carlo simulations, tricks can be used to deal with such as a limit~\cite{BloteDeng}. For our simulations using correspondence from such a partition-function approach, we have to take $\delta$ increasingly small to obtain the exact correspondence of the spectrum.

\begin{figure}
\includegraphics[width=0.5\textwidth]{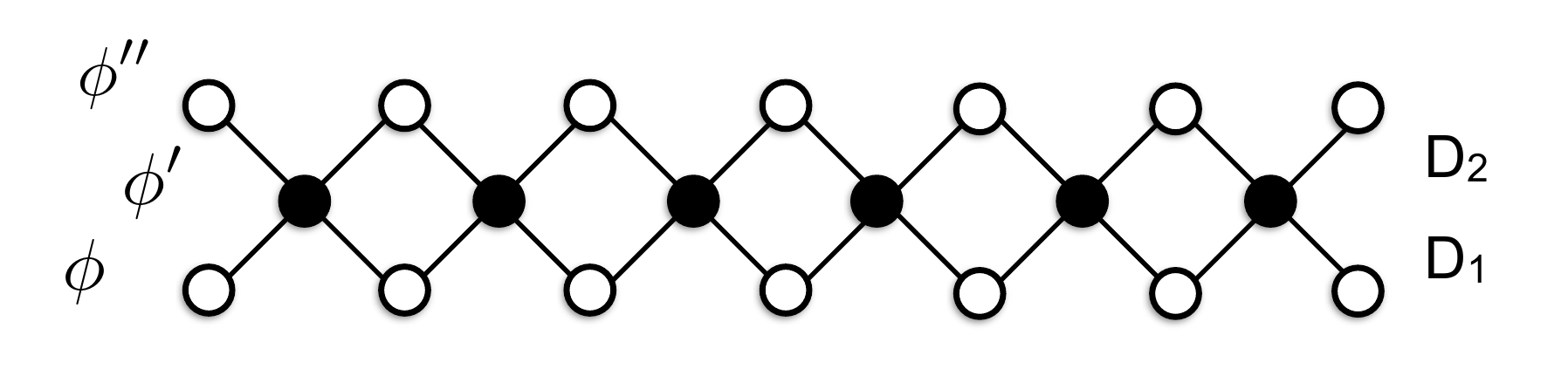}
\caption{The diagonal transfer matrix of square lattice. }  
\label{fig:diagonalT}
\end{figure}

Re-parametrizing the derived classical 2d anisotropic Ising model (see Fig.~\ref{fig:2dlattice}) we have
 \begin{align}
\label{eq:d+1C_Hamiltonian}
\beta H_c = -\sum_{\langle i,j \rangle} \left(K_x s^{[i,j]} s^{[i,j+1]} +  K_y s^{[i,j]} s^{[i+1,j]} \right),  
\end{align}
where $K_x , K_y$ are respectively the horizontal and vertical couplings,  $s^{[i,j]}=\pm1$ are classical spins at site $[i,j]$, and the sum runs over nearest neighbors on a square lattice. The classical canonical partition function of this model is given by
\begin{align}
Z_c  = \sum_{\{s \}} e^{\left(\sum_{\langle i,j \rangle} K_x s^{[i,j]} s^{[i,j+1]} +  K_y  s^{[i,j]} s^{[i+1,j]} \right)}.  
\label{eq:d+1C_partition}
\end{align}
Comparing Eq.~(\ref{eq:dQpartition}) with Eq.~(\ref{eq:d+1C_partition}) we then have the relations 
\begin{align}
K_x= J_{s}=  J_z \delta\tau, \quad K_y= J_{\tau} = \tanh^{-1}(e^{-2 \delta \tau J_x}),
\label{relations}
\end{align}
where we can set $J_z=1$ and $J_x=h$.  We thus obtain the relation between $h$ and $K_x,K_y$, 
\begin{align}
 \tanh K_y=  e^{-2 K_x h}.
\end{align}
The exact mapping is obtained in the limit 
  $K_x \to 0$ and $K_y\to 0$.

The case of a 3d classical Ising model on a cubic lattice, analogous to a 2d quantum Ising model in a transverse field on the square lattice, only introduces one more relation in additional to those Eq.~(\ref{relations}) for an extra coupling along a spatial direction. i.e., \begin{align}
& K_x= J_{s}=  J_z \delta\tau, \quad K_y= J_{s}=  J_z \delta\tau,   \notag \\
& K_z= J_{\tau} = \tanh^{-1}(e^{-2 \delta \tau J_x}). 
\label{2dQCrelations}
\end{align}
Such a $d$-dimensional quantum Ising model is mapped to a corresponding $(d+1)$-dimensional classical Ising model, which has homogeneous couplings along $d$ spatial dimensions, and is anisotropic in the extra (imaginary) temporal dimension.

\medskip\emph{\underline{(ii) Peschel's mapping in 2d:}}

In a work by Peschel~\cite{PeschelMap}, it was shown that a 2d classical Ising model with an isotropic coupling $K$ is in exact correspondence to a 1d quantum spin chain with Hamiltonian 
\begin{align}
\label{PeschelIsing}
H_q = -   \sum_{i=1}^{L-1} \sigma_x^{[i]}  -\delta \sigma_x^{[L]}   - \lambda \sum_{i=1}^{L-1} \sigma_z^{[i]}  \sigma_z^{[i+1]}, 
\end{align}
where $\delta = \cosh 2K$ and  $\lambda =\sinh^{2}K $, by using a transfer matrix technique. 
The transverse field labeled as $\delta$ at the right end can be neglected for large $L$. 
Then one arrives at the usual homogeneous chain.

Let us briefly review how this is derived. Consider the classical Hamiltonian of the 2d isotropic Ising model given by 
\begin{align}
\beta H_c = -\sum_{i,j} K (s^{[i,j]} s^{[i,j+1]} + s^{[i,j]} s^{[i+1,j]} ), 
\end{align}
where $s^{[i,j]} = \pm 1$ is a classical spin at site $[i,j]$ and $\beta$ is the inverse temperature. 
The partition function is given by 
\begin{align}
Z_c = \sum_{\{s\}} e^{ \left( K \sum_{i,j} (s^{[i,j]} s^{[i,j+1]} + s^{[i,j]} s^{[i+1,j]} ) \right)}. 
\end{align}
Firstly, by drawing the lattice diagonally (i.e., rotate the square lattice by 45 degrees), the sites can form a row as shown in Fig.~\ref{fig:diagonalT}, and these rows can be classified into two types: open circles and solid circles. This means that the number of rows must be even. 
Let now $N$ be the number of rows and $M$ is the number of sites in each row. 
Moreover, let $\phi_r$ denote all spins in row $r$ with $2^M$ possible values. 
In particular, the partition function can be represented by the diagonal-to-diagonal transfer matrix $W$ and $V$  as follows:
\begin{align}
Z_c = \sum_{\phi_1} \sum_{\phi_2}\cdots  \sum_{\phi_N} &(D_1)_{\phi_1,\phi_2}  (D_2)_{\phi_2,\phi_3}  (D_1)_{\phi_3,\phi_4} \notag \\ 
&\cdots(D_1)_{\phi_{N-1},\phi_N} (D_2)_{\phi_N,\phi_1}. 
\end{align}
Here, $(D_1)_{\phi_{j},\phi_{j+1}}$ contains all Boltzmann weight factors of the spins (from open circles to solid circles) in the adjacent rows $j$ and $j+1$. Similarly,  $(D_2)_{\phi_{j},\phi_{j+1}}$ contains the other type of spins (from solid circles to open circles). 
We now consider three rows labeled as $\phi, \phi', \phi''$, where $\phi = \{ s_1,s_2,...,s_M \}$ are the spins in the lower row and similarly for $\phi'$ and $\phi''$. Then the diagonal-to-diagonal transfer matrix is given by 
\begin{align}
& (D_1)_{\phi,\phi'} = e^{ K (  \sum_{j=1}^M (s_{j+1} s_j' + s_{j} s_j'  )  )   },  \notag \\
& (D_2)_{\phi',\phi''} = e^{ K (  \sum_{j=1}^M (s'_{j} s_j'' + s'_{j} s_{j+1}''  )  )   }. 
\end{align}
The partition function can thus be written as $Z_c = {\rm tr(}D_1D_2...D_1D_2 ) = {\rm tr}(D_1D_2)^{N/2} = {\rm tr} (V) ^{N/2}$. 
One can verify that $[H_q,V] = 0$ if the couplings are chosen to satisfy
\begin{align}
\delta = \cosh 2K, \quad \lambda =\sinh^{2}K. 
\end{align}
If the lattice size is large enough, then the single term $\sigma_L^x$ can be neglected. In this case the Hamiltonian can be written as a 1d quantum Ising chain with transverse field $h$, $H_q /\lambda= -   \sum_{i=1}^{L-1} h \sigma_x^{[i]}   -  \sum_{i=1}^{L-1} \sigma_z^{[i]}  \sigma_z^{[i+1]}$ with $h = 1/\lambda=1/\sinh^{2}K$. It is worth mentioning that the mapping is exact in the sense that no limit in any parameter needs to be taken (in contrast to, e.g., the partition-function approach, where we had $\delta\tau \rightarrow 0$).

\begin{figure}
\includegraphics[width=0.5\textwidth]{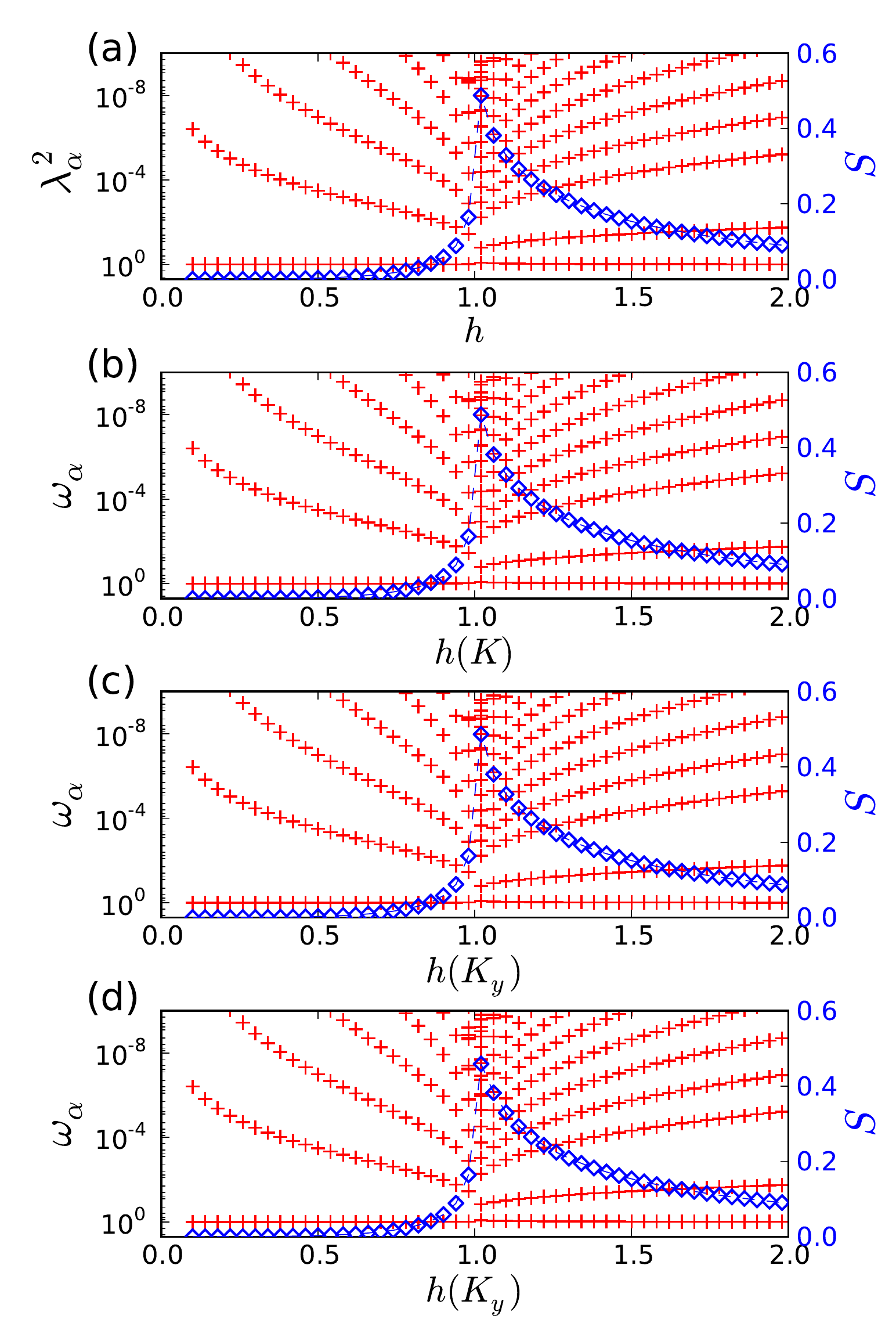}
\caption{[Color online]  (a) Entanglement spectra and entanglement entropy of the 1d quantum Ising model in a transverse field $h$ as obtained with iTEBD. (b,c,d) Corner spectra and corner entropy of: (b) the 2d classical isotropic Ising model, as a function of $h$, with isotropic coupling $K$ satisfying $1/h =\sinh^2K$; (c,d) 2d anisotropic classical Ising model with fixed $K_x=0.1$ (c), $K_x=0.01$ (d), and $K_y$ as a function of $h$ satisfying $\tanh K_y = e^{2 K_x h}$. The corner bond dimension is $\chi=20$ in all cases.}  
\label{fig:1dising_2dC_specn}
\end{figure}

\begin{figure}
\includegraphics[width=0.5\textwidth]{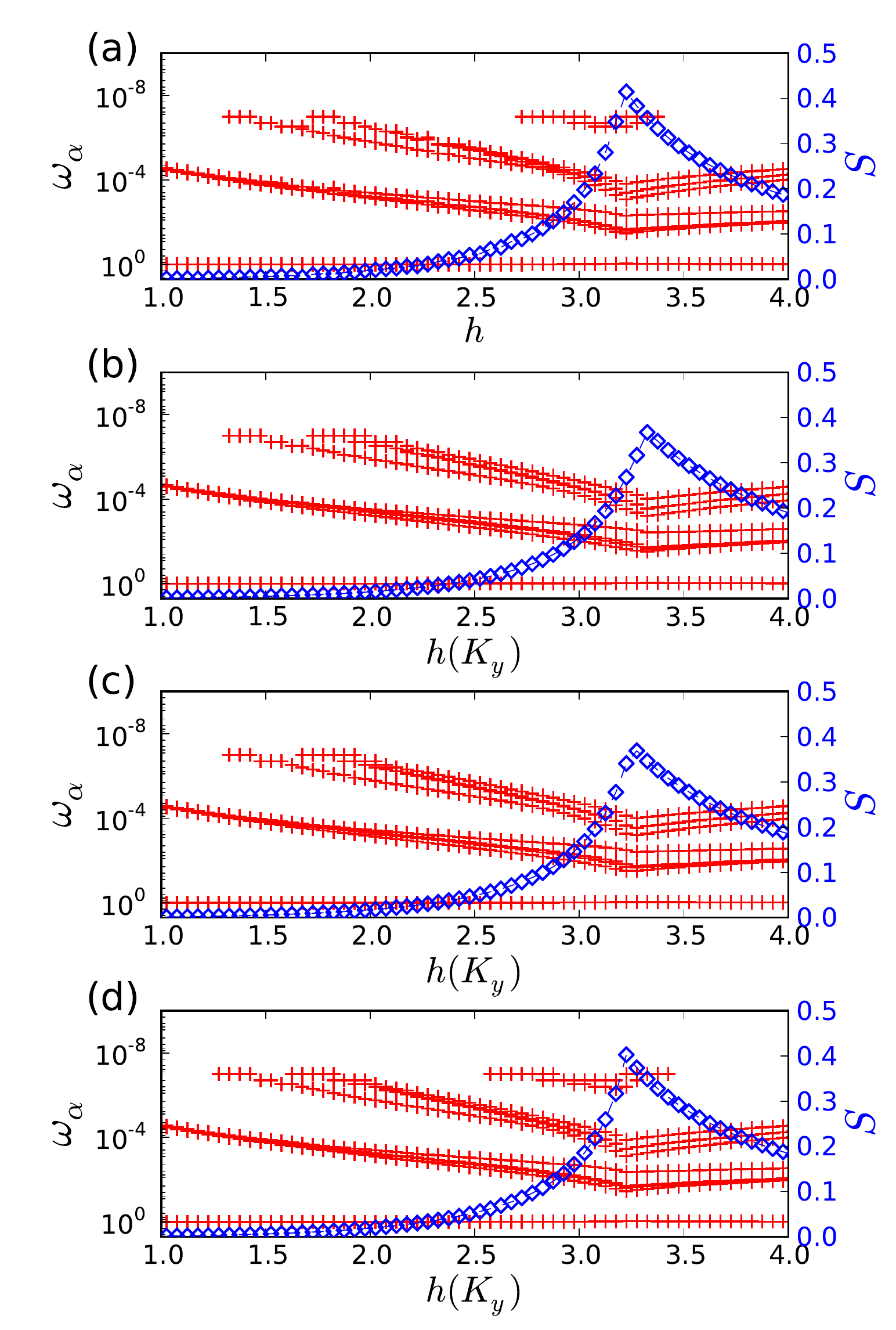}
\caption{[Color online] 
 Corner spectra and corner entropy of:
 (a) the 2d quantum Ising model in a transverse field $h$ by using the simplified one-directional 2d method \cite{3dCT};  (b,c,d) 3d anisotropic classical Ising model (also with the same method) with fixed  $K_x=K_y=0.1$ (b), $K_x=K_y=0.05$ (c), and $K_x=K_y=0.01$ (d), and $K_z$ as a function of $h$ satisfying $\tanh K_z = e^{2 K_x h}$. 
 The corner bond dimension for the CTs is $\chi = 4$ in all cases.}  
\label{fig:2dising_3d_spec}
\end{figure}

\medskip\emph{\underline{(iii) Numerical results:}} according to the mapping described above, we have computed the corner spectra $\omega_\alpha$ and the associated corner entropy for Ising models, first comparing the 1d quantum and 2d classical, and then the 2d quantum and 3d classical, using the numerical techniques mentioned earlier. On the one hand, the comparison of 1d quantum vs 2d classical is shown in Fig.~\ref{fig:1dising_2dC_specn}, where we also include in the second panel the mapping to the \emph{isotropic} classical Ising model by Peschel~\cite{PeschelMap}. Regarding the anisotropic classical model, the mapping becomes more and more precise as $\delta \tau \rightarrow 0$, i.e. as $K_x$ becomes smaller. In our results, when plotted with respect to the  same variables, we see a remarkably perfect agreement for all the numerical values of $\omega_\alpha$ and $S$ among all the models.  On the other hand, we show in Fig.~\ref{fig:2dising_3d_spec} our results comparing the 2d quantum vs 3d classical (anisotropic) case. The match in this case is not as perfect as in the 1d vs 2d case, but nevertheless, it is still quite remarkable, especially considering the inner workings and associated errors of the higher-dimensional numerical algorithms that we used.

\begin{figure} [t]
\includegraphics[width=0.5\textwidth]{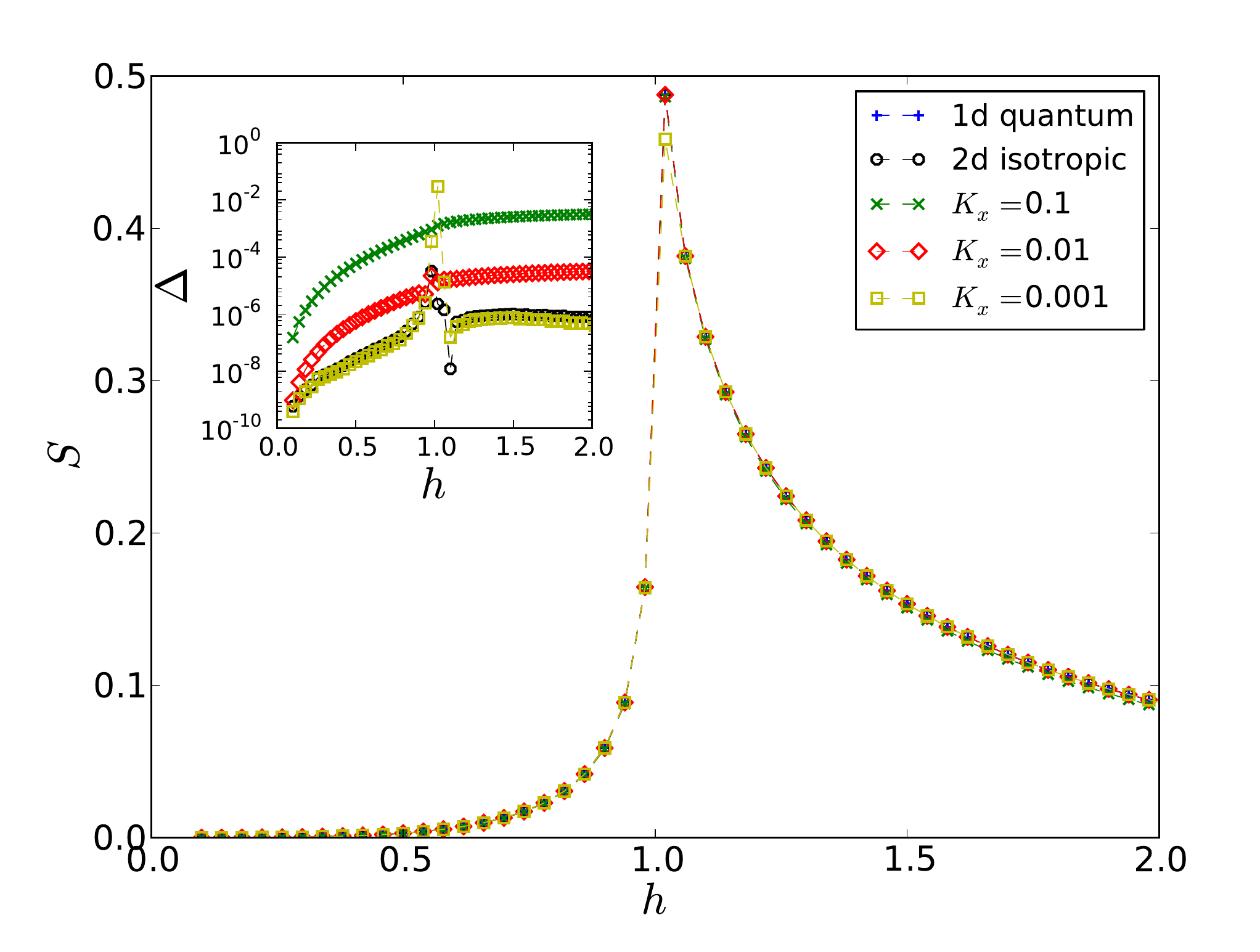}
\caption{[Color online]  Corner entropy of the 1d quantum Ising model, 2d classical isotropic Ising model  $(1/h =\sinh^2 K)$, and 2d classical anisotropic Ising model  with fixed $K_x=0.1$, $K_x=0.01$, and  $K_x=0.001$ $(\tanh K_y = e^{2 K_x h})$  as a function of the transverse field $h$ with bond dimension $\chi=20$. In the inset we show the difference $\Delta$ between the 2d corner entropies and the 1d entanglement entropy.}  
\label{fig:1dqising_2dcising_entropy}
\end{figure}

\begin {figure} [ht]
\includegraphics[width=0.5\textwidth]{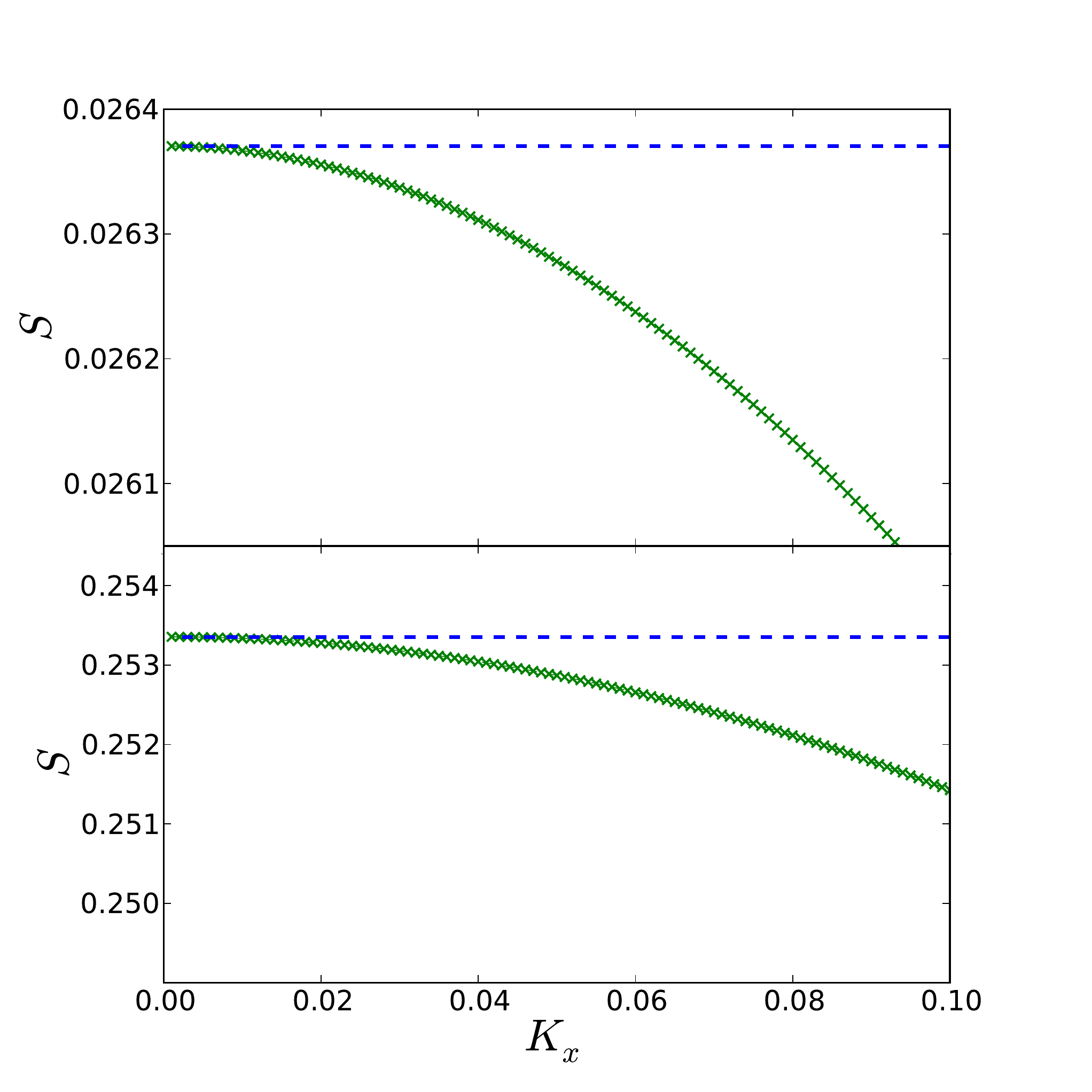}
\caption{[Color online] Corner entropy of the 2d classical anisotropic Ising model with fixed (upper) $h=0.8$ and (lower) $h=1.2$  as a function of $K_x$ with corner dimension $\chi=20$. The blue dashed lines show the entanglement entropy of the ground state of the corresponding 1d quantum Ising model obtained by using the iTEBD method.}
\label{fig:1dising}
\end{figure}

\begin{figure} [h]
\includegraphics[width=0.5\textwidth]{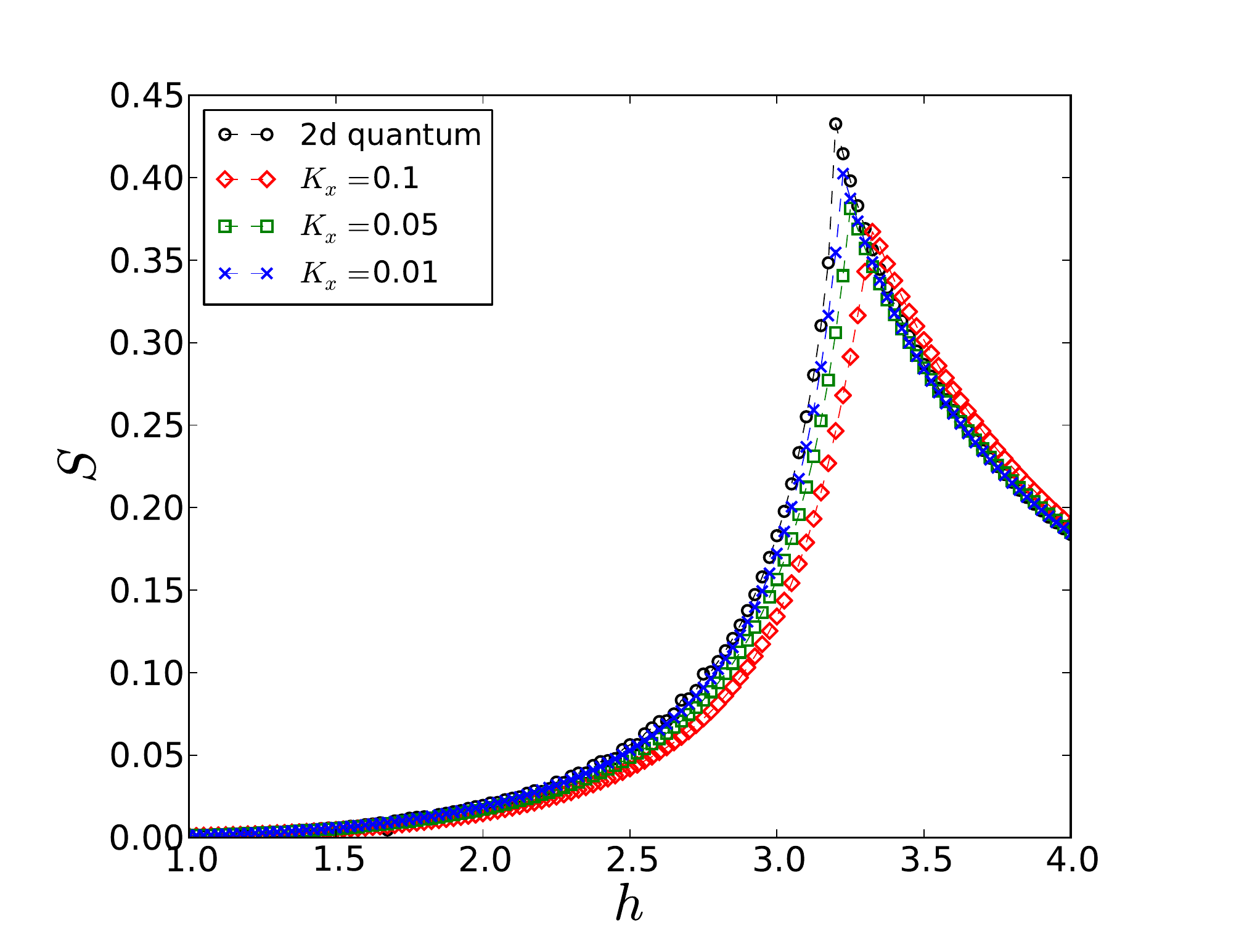}
\caption{[Color online]  Corner entropy of the 2d quantum Ising model and 3d classical anisotropic Ising model  with fixed $K_x=0.1$, $K_x=0.05$ , $K_x=0.01$ $(\tanh K_y = e^{2 K_x h})$  as a function of the transverse field $h$ with corner dimension $\chi=4$. }  
\label{fig:2dqisinf_3dcising_entropy}
\end{figure}

To understand further the data obtained from the corners, we show in Fig.~\ref{fig:1dqising_2dcising_entropy} the corner entropies in more detail, as well as difference between the 2d classical corner entropy and the one for the 1d quantum case (which equates the entanglement entropy). One can see in a more precise way that the entropy in the classical anisotropic case tends to the quantum one as the coupling $K_x$ tends to zero, as expected from Eq.~(\ref{relations}) for $\delta \tau \ll 1$. The effect of a finite $K_x$ is better appreciated in Fig.~\ref{fig:1dising}, where one can see clearly how the classical value tends to match as a limiting case the quantum value as $K_x \rightarrow 0$. Finally, in Fig.~\ref{fig:2dqisinf_3dcising_entropy} we show a comparison of the entropies for the 2d quantum vs 3d classical case. Again, as expected, the agreement between the quantum and the classical case improves as $K_x$ gets closer to zero. 

\subsubsection{Transverse field quantum N-Potts model in 1 dimension}

\emph{\underline{(i) Mapping:}} we now consider the 1d quantum N-state Potts model in 1d for $L$ sites. 
The corresponding 1d quantum Potts Hamiltonian is given by 
\begin{align}
\label{potts}
H_q 
& =- \sum_{i=1}^{L-1} \left( \sum_{n=1}^{N-1}  \left( Z^{[i] \dagger} Z^{[i+1]} \right)^n  \right)
    - h \sum_{i=1}^L \left(  \sum_{n=1}^{N-1} \left(X^{[i]} \right)^n \right)   \\
& = H_z+ H_x, \notag
\end{align}
where operators $Z$ and $X$ at every site satisfy 
\beq
Z \ket{q} = \omega^q \ket{q}, ~~~ X\ket{q} = \ket{q-1}, 
\eeq
with $\omega = e^{i 2 \pi / N}$ and $q \in \mathbb{Z}_N$. 

Similar to the case of the Ising model, the quantum canonical partition function is given again by 
\begin{align}
Z_q &= \tr \left(e^{-\beta  H_q} \right) = \sum_{\eta_z }  \Big{\langle} \{\eta_z \}  \Big{|} e^{-\beta H_q }   \Big{|}  \{\eta_z \}  \Big{\rangle}, 
\end{align}
but this time $ \Big{|} \{ \eta_z \} \Big{\rangle}  \equiv   |  \eta_z^{[1]},  \eta_z^{[2]}, \dots,    \eta_z^{[L]} \rangle $ is the diagonal basis of $Z$ for the $L$ spins, so that $\eta_z^{[i]} =  0,1,2,\dots,N-1, i = 1,2,...,L$. Proceeding as for the Ising model in the previous section, now we have a similar expression as in Eq.~(\ref{split}), but with $H_z$ and $H_x$ being the ones in Eq.~(\ref{potts}). For the Hamiltonian term $H_z$ we find 
\begin{align}
\label{eq:potts_Zterm}
& \Big{\langle}    \eta_z^{\prime [i]}   \eta_z^{\prime [i+1]}       \Big{|} 
  e^{\delta \tau  \left( \sum_{n=1} ^{N-1} (Z^{\dagger [i]} Z^{[i+1]})^n    \right) } 
\Big{|}  \eta_z^{[i]} \eta_z^{[i+1]}    \Big{\rangle}  \notag\\
& =  e^{\delta \tau  \vartheta_z } \delta_{ \eta_z^{[i]}  \eta_z^{\prime [i]}}  \delta_{ \eta_z^{[i+1]}  \eta_z^{\prime [i+1]}} , 
\end{align}
where  $\eta_z^{\prime [i]}  \equiv  \eta_z^{[i]} ({\tau+\delta \tau}) $ and   $\eta_z^{[i]}  \equiv  \eta_z^{[i]} ({\tau}) $. The coefficient  $ \vartheta_z$  is $ \vartheta_z=N-1$ if $ \eta_z^{[i]}  =  \eta_z^{[i+1]}$, and $\vartheta_z=-1$ otherwise. Additionally, for the term $H_x$ one has 
\begin{align}
\label{eq:potts_Xterm}
& \Big{\langle}    \eta_z^{\prime [i]}     \Big{|}  e^{\delta \tau  h ( \sum_{n=1}^{N-1} ( X^{[i]})^n  )  }  \Big{|}  \eta_z^{[i]}     \Big{\rangle}  \notag\\
& = \Big{\langle}  \eta_z^{\prime [i]}  \Big{|}  \cosh(\delta \tau  h)  \mathbb{I} + \sinh(\delta \tau  h)  ( \sum_{n=1}^{N-1} ( X^{[i]})^n  )  \Big{|}  \eta_z^{[i]}     \Big{\rangle}  \notag\\
&  =
  \begin{cases}
    \cosh(\delta \tau  h)      & \quad \text{if } \eta_z^{[i]}  =  \eta_z^{\prime [i]}  \\
    \sinh(\delta \tau  h)  & \quad \text{otherwise}.  \\
  \end{cases}
\end{align}

\begin{figure*}
\includegraphics[width=1.0\textwidth]{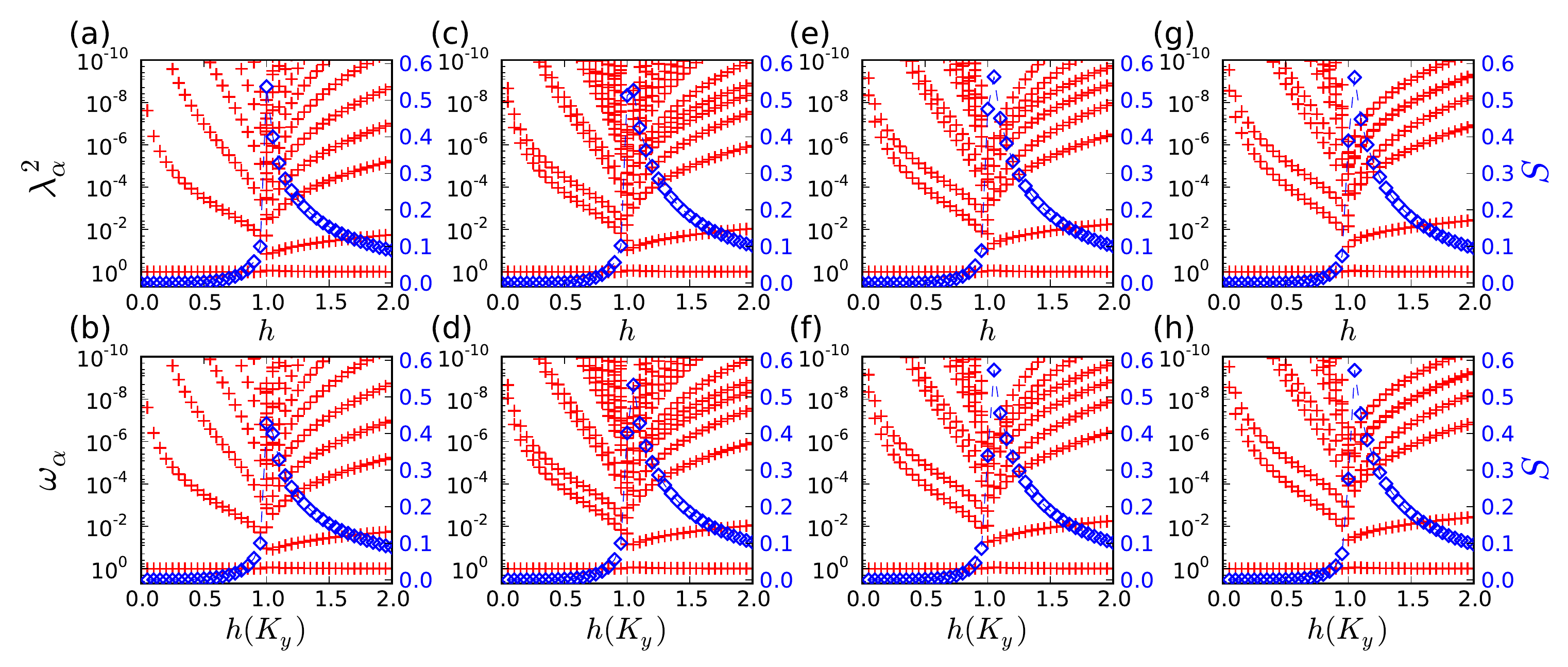}
\caption{[Color online] Entanglement spectra and entanglement entropy for the 1d quantum N-state Potts model in transverse field $h$ for (a) N=2,  (c) N=3,  (e) N=4, and  (g) N=5 by using iTEBD method. Corner spectra and corner entropy for the 2d classical N-state Potts model as a function of $h$, where $h$ is a function of $K_y$ as in Eq.~(\ref{QCpotts}), with $K_x=0.01$  for (b) N=2,  (d) N=3,  (f) N=4, and (h) N=5 computed with the 2d directional CTM method. The corner bond dimension is $\chi = 20$ in all cases.}  
\label{fig:1d_2d_Potts}
\end{figure*}

For the classical case, the Hamiltonian of the 2d classical N-state Potts model on a square lattice is defined by 
\begin{align}
\label{eq:classical_potts}
\beta H_c = -\sum_{\langle i,j \rangle} \left( K_x \delta_{s^{[i,j]},s^{[i,j+1]} } +  K_y \delta_{s^{[i,j]},s^{[i+1,j]} } ) \right),   
\end{align}
with ``Potts spin variables" $s^{[i,j]}=0,1,2,...,N-1$ at each site. 
The classical partition function is then 
\begin{align}
\label{eq:classical_potts_partition}
Z_c =\sum_{ \{ s \} }   e^{\left( \sum_{\langle i,j \rangle } K_x  \delta_{s^{[i,j]},s^{[i,j+1]} }  
                                                                                           + K_y \delta_{s^{[i,j]},s^{[i+1,j]} } \right)}. 
\end{align}
From Eqs.~(\ref{eq:potts_Zterm}), ~(\ref{eq:potts_Xterm}), and (\ref{eq:classical_potts_partition}) one finds the relations 
\begin{align}
\label{QCpotts}
K_x = N \delta \tau, ~~ \tanh(\delta \tau  h ) = e^{-K_y}, 
\end{align}
which establish the quantum-classical mapping.

\medskip\emph{\underline{(ii) Numerical results:}} as we did for the case of the Ising model, now we have benchmarked  the quantum-classical correspondence by computing numerically the corner spectra $\omega_\alpha$ and their associated corner entropy for several quantum and classical Potts models. Our results are summarized in Fig.~\ref{fig:1d_2d_Potts}, where we show the corner spectra and corner entropy for the 1d quantum and 2d classical $N$-state Potts models for $N$=2,3,4 and 5. Again, we find a remarkable almost-perfect match for the corner properties as computed with different methods for 1d quantum and 2d classical systems, once the parameters in the models are rescaled according to the relations found in the previous section. The spectrums for the 2-state Potts model coincide with those of the Ising model, as expected. As $N$ increases, we find small variations in the corner for different values of $N$, even though the branches corresponding to the lowest corner spectra seem to be very similar for all the computed $N$.

\subsection{Suzuki's approach for the quantum XY model} 

\begin{figure}[h]
\includegraphics[width=0.4\textwidth]{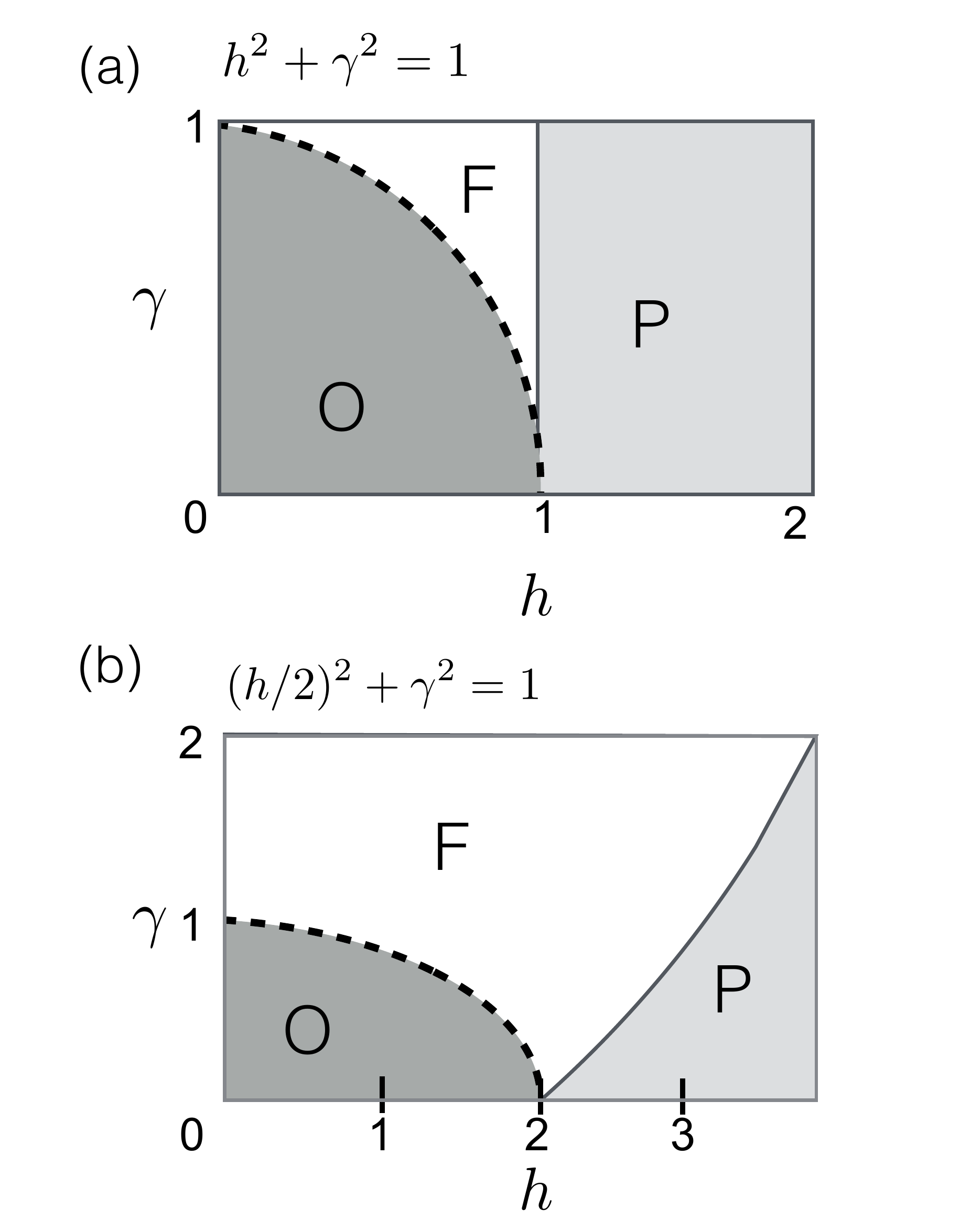}
\caption{[Color online] Phase diagram of (a) 1d  \cite{XYmodel}  and (b) 2d  \cite{2dXYmodel} quantum XY model. There are three phases: oscillatory (O), ferromagnetic (F), and paramagnetic (P). The equation on top is the Barouch-McCoy circle~\cite{BarouchMcCoy} that sets the boundary between the oscillatory and non-oscillatory ferromagnetic regions (which is only a crossover). The separation between F and P in (a) is at $h=1$ and in (b) the exact location is not known and only indicated schematically.}  
\label{fig:pd_1dXY}
\end{figure}

In a work by Suzuki \cite{SuzukiXY} it was proven that a 2d classical Ising model in the absence of a magnetic field and with anisotropic couplings is ``equivalent", in the sense of having the same expectation values and physical properties, to the ground state of a XY quantum spin chain. Unlike the partition function approach, which maps a quantum model to a classical model in one dimension higher, Suzuki's approach works from the other direction: it starts from the $(d+1)$-dimensional classical partition function, and then builds a $d$-dimensional quantum model with the same physics. We note that the mapping is exact and does not involve the limit. However, if one uses the quantum XY model to study the transverse-field Ising model, then a similar limit  needs to be taken. Morever, it was known that there is a range of couplings in the quantum XY model that there is no valid classical correspondence (see the ``O'' region in Fig.~\ref{fig:pd_1dXY}).

\medskip\emph{\underline{(i) The mapping:}} let us review the theory behind this approach by considering first the classical Hamiltonian of the anisotropic 2d XY model, i.e., 
\begin{align}
\beta H_c = -\sum_{i,j} \left( K_x s^{[i,j]} s^{[i,j+1]} +  K_y s^{[i,j]} s^{[i+1,j]} \right), 
\end{align}
where indices $i,j$ denote respectively rows and columns, $K_x, K_y$ are the horizontal and vertical couplings, $s^{[i,j]} = \pm 1$ are classical spin variables at each site, and $\beta$ is the inverse temperature. For concreteness let us imagine that we have a finite periodic square lattice with $N\times M$ sites.  

The canonical partition function is given by
\begin{align}
Z_c  = \sum_{\{s \}} e^{\left(K_x \sum_{i,j}  s^{[i,j]} s^{[i,j+1]} +  K_y \sum_{i,j}  s^{[i,j]} s^{[i+1,j]} \right)}.  
\end{align}
The first sum inside the brackets in the exponential is over horizontal edges, and the second over the vertical ones. Let $N$ be the number of rows in the lattice and $M$ the number of sites in each row. Now let $\phi_r$ denote all spins in row $r$, so that $\phi_r$ has $2^M$ possible values. 
The partition function can thus be thought of as a function of $\phi_1$,.....,$\phi_N$, and can be rewritten as 
\begin{align}
Z_c  = \sum_{\phi_1}...\sum_{\phi_N}  T_{ {\phi_1},{\phi_2}}... T_{{\phi_{N-1}},{\phi_N}} T_{{\phi_N}.{\phi_1}}, 
\end{align}
Here $T_{\phi_i ,\phi_{i+1} }$ is the 1d transfer matrix of the system, which contains all the Boltzmann weight factors of the spin in the adjacent rows. 

Let $\phi= \{ s_{1}s_{2},...s_{M} \}$ be the spins in a given row,  and  $\phi' = \{ s'_{1},s'_{2},...s'_{M} \}$ the ones in the following row. Then the transfer matrix is given by  
\begin{align}
T_{\phi,\phi' } 
&= e^{ \left(K_x \sum_{i}  s_{i} s_{i+1}  +K_y\sum_{i} s_{i} s'_{i} \right) } \notag \\
& = e^{K_x \sum_{i}  s_{i} s_{i+1}}\times e^{K_y\sum_{i} s_{i} s'_{i}} \notag \\
& \equiv V_1 V_2. 
\end{align}
Here $V_1$ can be decomposed as a product of   $2 \times 2$ matrices,  
\begin{align}
(V_1)_{s_i,s_{i+1}} = 
\begin{pmatrix}
e^{K_x} &   e^{-K_x} \\
e^{-K_x} &   e^{K_x} 
\end{pmatrix}, 
\end{align}

\begin{figure*}
\includegraphics[width=1.0\textwidth]{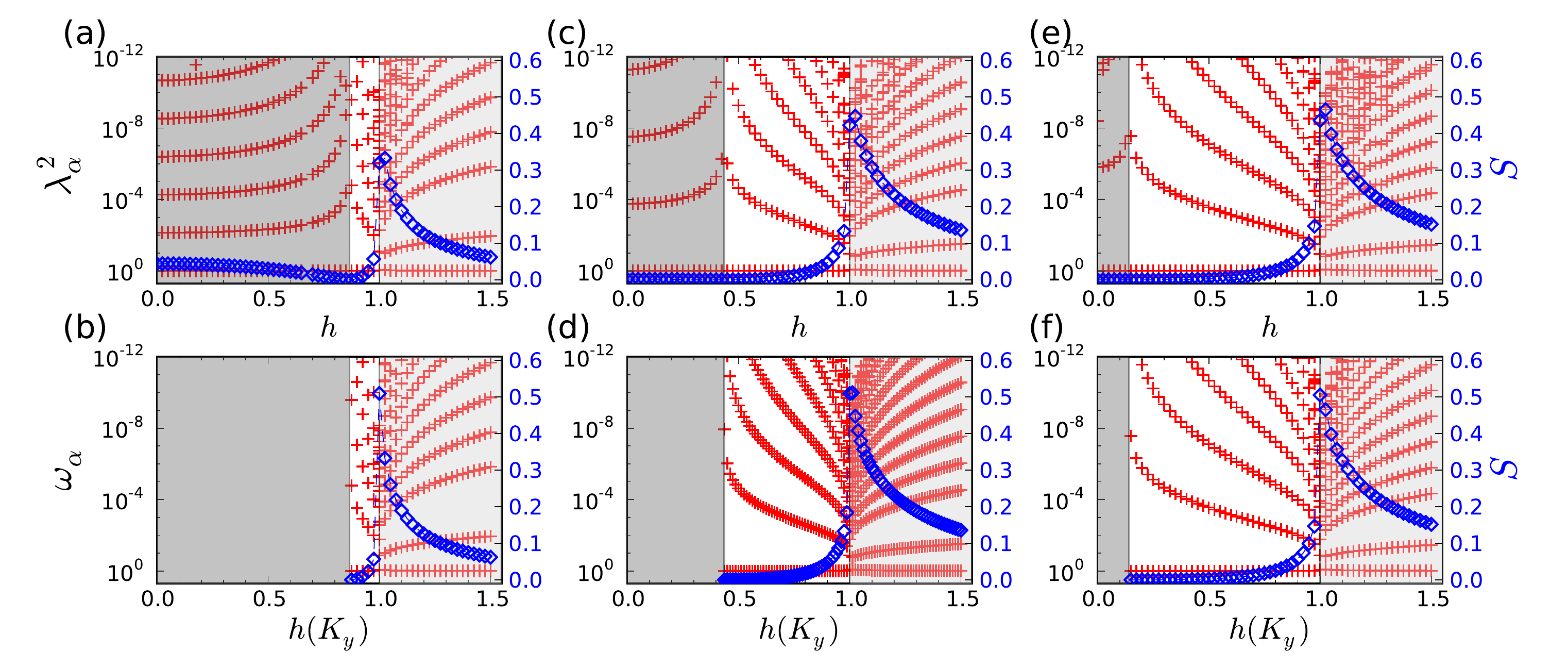}
\caption{[Color online]  Corner spectra and entropy for the 1d XY quantum spin chain, with (a) $\gamma = 0.5$, (c) $\gamma = 0.9$, and (e) $\gamma = 0.99$, together with the corresponding anisotropic 2d classical Ising model (b), (d), (f), respectively, with $K_x$ and $K_y$ being functions of $h$ as described in Eq.(\ref{rela}). The corner bond dimension (equivalent to the MPS bond dimension in the 1d quantum case) is $\chi = 40$ in both cases. The correspondence of parameters has a solution only for values of $h$ larger than (b)  $h \approx 0.85$, (d) $h \approx 0.4$, (f)  $h \approx 0.1$, and therefore the left hand side of each plot in the lower panel is empty.}  
\label{fig:1dQXY_gamma}
\end{figure*}

which can also be written as 
\begin{align}
 (V_1)_{s_i,s_{i+1}} &= e^{K_x}\mathbb{I}  +  e^{-K_x}\sigma_x  \notag \\
 &= e^{K_x} ( \mathbb{I} + e^{-2K_x}\sigma_x  )  \notag \\
 & = (2\sinh2K_x  )^{1/2} e^{K_x^*\sigma_x } \equiv V_1(i),
\end{align}
with $\mathbb{I}$� the $2 \times 2$ identity matrix, $\sigma_x$ the x-Pauli matrix, and where we define $\tanh K_x^* \equiv e^{-2K_x} $ (and $\tanh K_x \equiv e^{-2K_x^*}$), as well as use the relation $\sinh 2K_x \sinh 2K_x^* = 1 $. Moreover, one has the $4 \times 4$ matrix $  \big(  (V_2)_{s_i,s_j;s'_i,s'_j} \big)  \delta_{s_i,s'_i}  \delta_{s_j,s'_j}  $ given by 
\begin{align}
& (V_2)_{s_i,s_j;s'_i,s'_j}  \notag\\
 &= \bordermatrix{~  
                                       &  (+1,+1)  &  (+1,-1)  & (-1,+1)  & (-1,-1) \cr
                     &  e^{K_y} &  0&  0&  0\cr
                   &  0 &   e^{-K_y} &  0&  0\cr
                  & 0 &   0 &   e^{-K_y}  &  0\cr
                 &  0&   0&  0 &   e^{K_y} \cr                  
                  } \notag\\
                  &=  \exp (K_y \sigma_z^i \sigma_z^{i+1} )  \notag\\
                  & =  \cosh K_y  \mathbb{I}^{i} \mathbb{I}^{i+1}+ \sinh K_y \sigma_z^i \sigma_z^{i+1}   \notag\\
                    &   \equiv  V_2(i,i+1) .  
\end{align}

It is clear that the partition function is the trace of a matrix product, given by 
\begin{align}
Z_c = \tr (V_1V_2...V_1V_2 ) =\tr (V_1V_2)^N. 
\end{align}
Thus, $Z_c$ can also be written as 
\begin{align}
Z_c = \tr (V_2^{1/2} V_1V_2^{1/2} )^N =\tr (V)^N, 
\end{align}
or 
\begin{align}
Z_c  = \tr (V_1^{1/2} V_2V_1^{1/2} ) =\tr (V')^N, 
\end{align}
where 
\begin{align}
V_1 =  (2\sinh2K_x  )^{M/2} e^{(K_x^* \sum_{i=1}^m \sigma_z^i )}, 
\end{align}
and 
\begin{align}
V_2=  e^{(K_y  \sum_{i=1}^M  \sigma_z^i \sigma_z^{i+1} )}. 
\end{align}

The next step is to show that $V$ and the quantum Hamiltonian $H_q$ for the 1d quantum XY model can commute, and therefore have common eigenvectors. The usual XY quantum spin chain is defined by the Hamiltonian
\begin{align}
H_q = - \sum_i \left( J_x \sigma_x^{[i]}  \sigma_x^{[i+1]} +J_y \sigma_y^{[i]}  \sigma_y^{[i+1]} \right) +h \sum_{i} \sigma_z^{[i]}, 
\end{align}
where $\gamma = (J_x-J_y)$ is the anisotropy, and $h$ the magnetic field. 
The phase diagram of the model is well known \cite{XYmodel} and is sketched in Fig.~\ref{fig:pd_1dXY}a.

To prove that the commutator of $V$ and $H_q$ can sometimes be zero, we first define $V_2(i,i+1)^{\frac{1}{2}} \equiv  v_2(i,i+1) $.  The 1d quantum Hamiltonian is a sum of two-body operators $H_q= \sum_i h(i,i+1)$. Thus, the commutator reads \begin{align}
[V, H_q] &=  \sum_i [V, h(i,i+1)  ] \notag \\
& = \sum_i  \big(\cdots[ v_2(i-1,i)v_2(i,i+1) v_2(i+1,i+2)\notag \\ 
&  V_1(i)  V_1(i+1) v_2(i-1,i)v_2(i,i+1)v_2(i+1,i+2)   \notag \\ 
& , h(i,i+1) ]\cdots \big)  =0.
\end{align}
The last equality imposes a constraint on the couplings of the classical and quantum models in order for the commutator to vanish. One can see that this implies the relations between the couplings
\begin{align}
& \frac{J_y}{J_x} = e^{-4 K_x},~~  \frac{h}{J_x}=  2e^{-2 K_x} \coth(2K_y),  
\label{rela}
\end{align}
which make explicit the quantum-classical mapping. Importantly, for fixed $h, J_x$ and $J_y$, these equations do not have a real solution in the oscillatory phase of Fig.~\ref{fig:pd_1dXY}, so that the mapping is only valid outside of that phase. Finally, the mapping can also be extended easily to the 3d classical vs 2d quantum case,  by  considering a 2d homogeneous coupling $K_x = K_y = K$ and adding an extra equation for $K_z$,  i.e.,
\begin{align}
& \frac{J_y}{J_x} = e^{-4 K},~~ \frac{h}{J_x}=  2e^{-2 K} \coth(2K_z). 
\end{align}

\begin{figure}[h]
\includegraphics[width=0.5\textwidth]{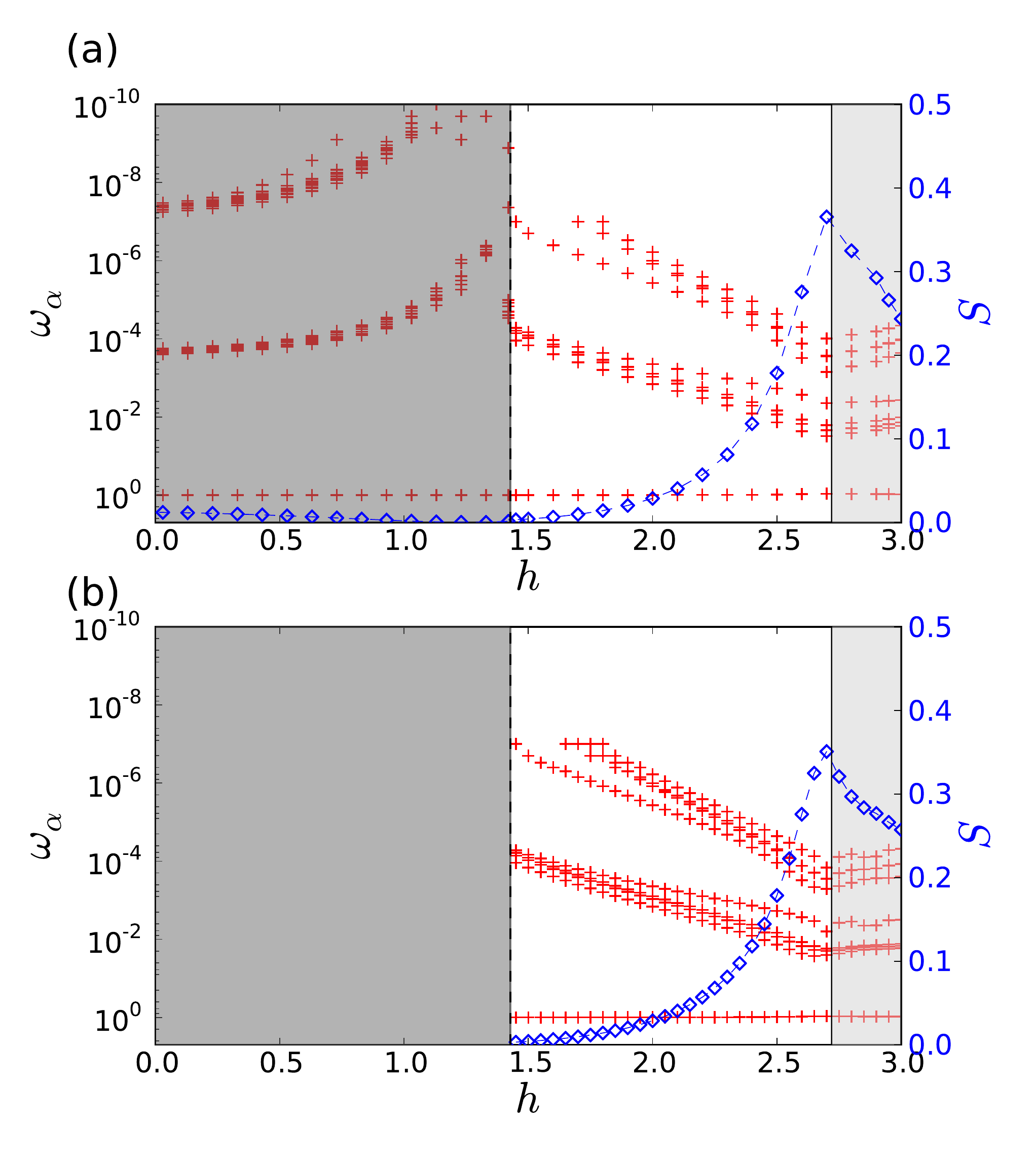}
\caption{[Color online]   Corner spectra and corner entropy of: (a) 2d quantum XY model with $\gamma = 0.7$ in a transverse field $h$ by using the simplified one-directional 2d method \cite{3dCT}; (b) the corresponding  3d anisotropic classical Ising model as a function of $h$ satisfying Eq.~(\ref{rela}). The corner bond dimension is $\chi=4$ in all cases. The correspondence of parameters has a solution only for values of $h$ larger than $h \approx 1.45$,  and therefore the left hand side of the plot in the lower panel is empty.}  
\label{fig:2dqXY_3dcising07}
\end{figure}

\medskip\emph{\underline{(ii) Numerical results:}} we have explicitly checked this equivalence by computing numerically the corner spectra and the associated corner entropy for the quantum and classical XY models in 1d, 2d and 3d. For the 1d quantum vs 2d classical case, this is shown in Fig.~\ref{fig:1dQXY_gamma} for different values of the anisotropy in the quantum XY model. The expressions in Eq.~(\ref{rela}) have only a real solution for $K_x, K_y$ if the value of $h$ is outside of the oscillatory phase,  as shown in the plots. We can see that the agreement between the quantum and classical corner spectra and corner entropy is remarkably good, both qualitatively and quantitatively, with a slightly larger error around the critical region $h = 1$. 
The comparison between 2d quantum vs 3d classical can be found in Fig.~\ref{fig:2dqXY_3dcising07}. 
Again in this case the match between the numerically-computed classical and quantum values is quite remarkable, considering the different numerical techniques that were used in this case. 

\section{2d corner phase transitions}
\label{Sec6} 

We now show how the study of corner properties can provide other useful information when studying a quantum or classical many-body system. In particular, we show how the corner spectra and corner entropy from 2d rCTMs (i.e., the CTMs obtained from the 2d TN for the norm) are useful in determining phase transitions without the need to compute physical observables. 

The usual way to study quantum and classical phase transitions is through the study of observables, which have specific properties at the transition point (e.g., the singular behavior of the observable). The study of entanglement and correlations in many-body systems has shown us that it is actually possible to study these transitions from properties of the state only,  such as entanglement entropy, fidelities \cite{fidelity}, entanglement spectra \cite{eHam}, and similar quantities. Following this trend, in this section we show that one can assess phase transitions from properties of the corners only, in particular the rCTM that we introduced in Sec.~\ref{sec:intro}. This is very useful in the context of numerical simulations of, e.g., 2d quantum many-body systems, since such corner objects are produced ``for free" (e.g., in the infinite-PEPS method with a full or fast-full update \cite{iPEPS, ffUpdate}). In what follows we show three practical examples where phase transitions, both topological and non-topological, can be clearly pinpointed by looking only at the corner objects. 

\subsection{2d quantum XXZ model}

First we consider the 2d quantum XXZ model for spin-1/2 on an infinite square lattice, under the effect of a uniform magnetic field $h$ along the z-axis. Its Hamiltonian is given by 
\begin{align}
H_q = -\sum_{\langle i,j \rangle} \left(  \sigma_x^{[i]} \sigma_x^{[j]} + \sigma_y^{[i]} \sigma_y^{[j]}  -\Delta \sigma_z^{[i]} \sigma_z^{[j]} \right) - h\sum_i \sigma_z^{[i]}, 
\end{align} 
where as usual the sum $\langle i,j \rangle$ runs over nearest neighbors on the 2d square lattice, and $\Delta$ is the anisotropy. In the large $\Delta>1$ limit, it has been shown \cite{xxz} that a first-order transition takes place at some point $h_1$ from a N\'{e}el phase to a spin-flipping phase. As the field increases further, another phase transition at $h_2 = 2(1+\Delta)$ occurs towards the fully polarized phase.

\begin{figure}
\includegraphics[width=0.5\textwidth]{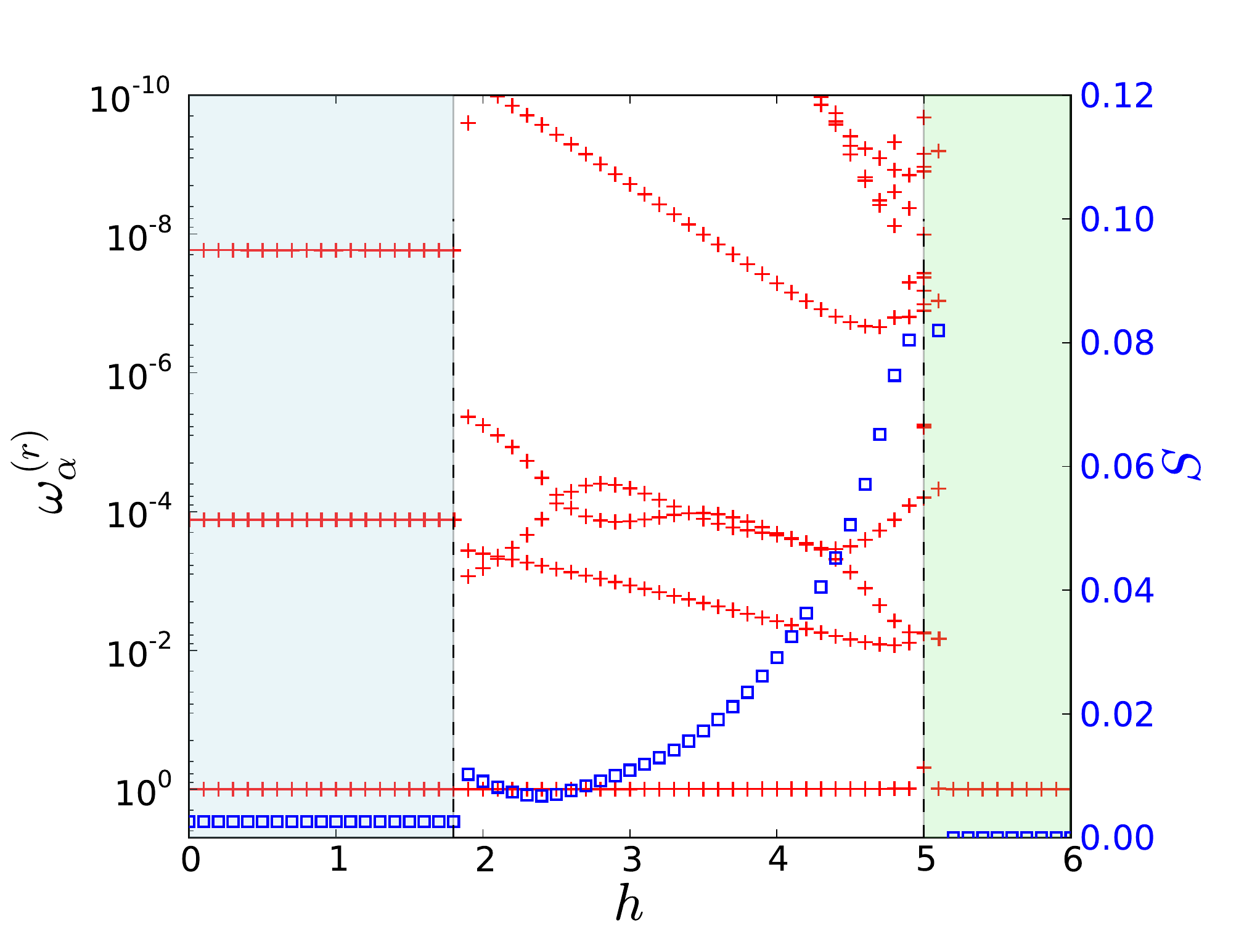}
\caption{[Color online]  Corner spectra $\omega^{(r)}_\alpha$ for the norm of the numerical $D=2$ PEPS for the XXZ model in a field, at $\Delta = 1.5$, on the square lattice with $\chi=40$, together with the corner entropy computed from the corner spectra. }  
\label{fig:2DXXZ}
\end{figure}

Here we consider the case with $\Delta= 1.5$. We have approximated the ground state of the model using the iPEPS algorithm with simple update and bond dimension $D=2$ \cite{su}, and then computed the reduced corner spectra $\omega^{(r)}_\alpha$ and entropy of the double-layer tensor defining the norm via the directional CTM approach, as a function of $h$. Our results are shown in Fig.~\ref{fig:2DXXZ}, where one can clearly see that the two phase transitions are clearly pinpointed by the spectrum and the entropy. In particular, we observe the first transition happening at $h_1 \approx 1.8$, and the second one at $h_2=5.0$. 

\subsection{Perturbed $\mathbb{Z}_N$ topological order}

Here we consider exact wavefunctions that exhibit topological phase transitions for $\mathbb{Z}_2$ and $\mathbb{Z}_3$ topological order. 
 
\begin{figure}[h]
\includegraphics[width=0.5\textwidth]{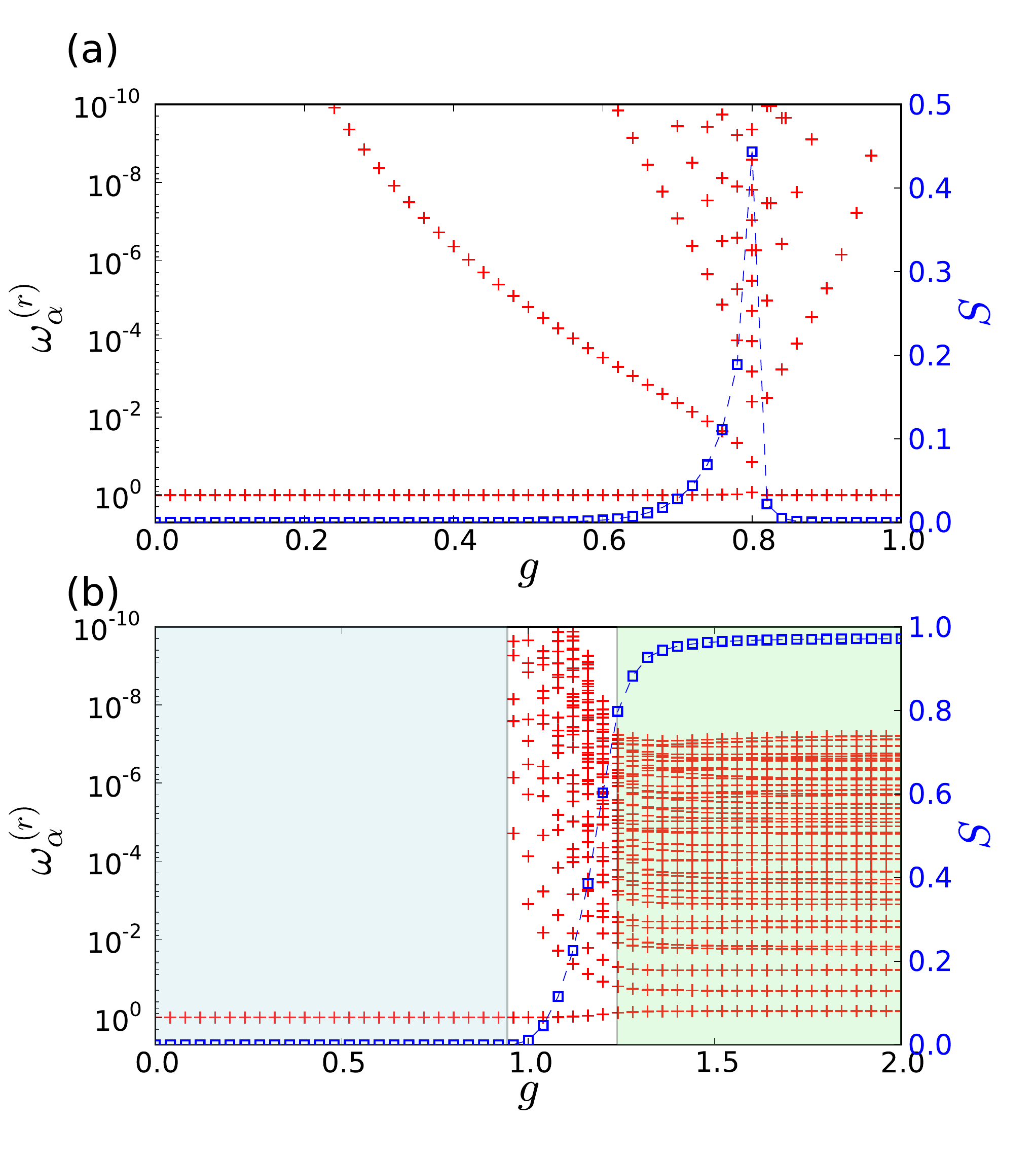}
\caption{[Color online] Corner spectra and corner entropy of the (a) $\mathbb{Z}_2$ and (b) $\mathbb{Z}_3$ topological PEPS with perturbation $g$ on the square lattice with CTM bond dimension $\chi=20$. In (b) the lines show the first transition point at $g_1 \approx 0.944$ as well as the second transition point at $g_2 \approx 1.238$ \cite{Z3_2}.}  
\label{fig:ZNTO}
\end{figure}

\medskip\underline{\emph{(i) 2d perturbed $\mathbb{Z}_2$ Toric Code PEPS:}} 
we consider the 2d PEPS on a square lattice for the Toric Code ground state \cite{TC, IsingPEPS}, perturbed by a string tension $g$. This can be represented by a tensor $A_{\alpha \beta \gamma \delta}^{i,j,k,l}$ with with four physical indices $i,j,k,l=0,1$  and four virtual indices $\alpha,\beta, \gamma,\delta=0,1$. The coefficients of the tensor are given by 
\begin{align}
 A_{i,j,k,l}^{i,j,k,l} = \left\{ 
  \begin{array}{l l}
    g^{i+j+k+l}, & \quad \text{if $i+j+k+l=0$ mod 2}, \\
    0, & \quad \text{otherwise}.
  \end{array} \right.
\end{align} 
The norm of this state can be described by a double-layer 2d TN on a square lattice, where at every site one has the tensor $\mathbb{T}_{ijkl}^{ijkl} \equiv \mathbb{T}[ijkl]$, with coefficients
\begin{align}
& \mathbb{T}[0000] =1, \quad  \mathbb{T}[1111]= g^8, \notag \\
&\mathbb{T}[0011] =\mathbb{T}[0110]= \mathbb{T}[1100]=\mathbb{T}[1001] = g^4\notag \\
&\mathbb{T}[0101] = \mathbb{T}[1010] =g^4.
\label{db}
\end{align} 
Parameter $g$ is used to tune a crossover from a topological to a trivial phase. For $g=1$ the state reduces to the ground state of the Toric Code model with  $\mathbb{Z}_2$ topological order. 
For $g=0$ it reduces to the polarized state $\ket{0, 0, \cdots, 0}$. There is a quantum phase transition between these two phases which, as shown in Ref.~\cite{Z2_deform}, occurs at  $g_c\approx 0.802243$. 
One can see, moreover, that the double tensor $\mathbb{T}$ consists of two copies of the partition function of the 2d classical Ising model in Eq.~(\ref{cIsing}). In fact, one also finds the relation $g=(\sinh(\beta))^{1/4} $, with $g$ the perturbation parameter of the Toric Code and $\beta$ the inverse temperature of the Ising model. Both models, therefore, belong to the same universality class. In this case we have implemented the directional CTM method on the norm tensor $\mathbb{T}$ \cite{dirCTM} to study the corner properties. 
This is shown in  Fig.~\ref{fig:ZNTO}a, where one can see that the corner spectrum and its associated entropy clearly pinpoint the quantum phase transition.  

\medskip\underline{\emph{(ii) 2d perturbed $\mathbb{Z}_3$ topological order:}} furthermore, we consider a 2d PEPS with $\mathbb{Z}_3$  topological order under perturbations described by deformations $\{q_0,q_1,q_2\}$. The PEPS is given by a tensor $A_{\alpha,\beta, \gamma,\delta}^{i,j,k,l}$ with with four physical indices $i,j,k,l=0,1,2$  and four virtual indices $\alpha,\beta, \gamma,\delta=0,1,2$, with coefficients  
\begin{align}
 A_{i,j,k,l}^{i,j,k,l} = \left\{ 
  \begin{array}{l l}
    q_0^{n_0} q_1^{n_1} q_2^{n_2}, & \quad \text{if $i+j+k+l=0$ mod 3}, \\
    0, & \quad \text{otherwise}, 
  \end{array} \right.
\end{align} 
where $n_0,n_1,n_2$ means the number of the inner indices in 0, 1, and 2 respectively. 
We first study the case $q_0=1, n_1=0, q_2=g$. In such a case, the bond indices of the wavefunction live in an effective 2d Hilbert space spanned by $|0\rangle$ and $|2\rangle$. At $g=0$, the remaining tensor represents a product state of all state $0$. Therefore, the region near $g=0$ is a  trivial phase that is adiabatically connected to a product state. At $g> 0$, the nonzero components of the double-layer tensor $\mathbb{T}$ for the norm are
\begin{align}
 &\mathbb{T}[0222] =   \mathbb{T}[2022] = \mathbb{T}[2202]=  \mathbb{T}[2220]= g^6 \notag \\
 & \mathbb{T}[0000]= 1, 
\end{align} 
where we used the same notation as in Eq.~(\ref{db}). For $g \gg 1$ one can neglect the component $\mathbb{T}[0000]$, and the tensor becomes mathematically equivalent to the one for the classical dimer model at Rokhsar-Kivelson (RK) point, which is critical \cite{RK}, and where the topological degenerate ground state is an  equal weight superposition of all possible configurations in a given winding parity sector on the square lattice. It was shown in Ref.~\cite{Z3_2} that for $0.944 \leq  g < 1.238$ the PEPS belongs to the $\mathbb{Z}_3$ topologically ordered phase \cite{Z3_1}, whereas for $g > 1.238$ the state is critical. 

We have computed the rCTM spectra obtained by contracting the TN for the norm using the directional CTM approach \cite{dirCTM}, and as a function of the deformation $g$. This is shown in  Fig.~\ref{fig:ZNTO}b. The corner spectra show different patterns depending on the phase: in the trivial phase only one eigenvalue is non-zero, whereas more eigenvalues become populated in the topological and critical phases. The two transitions are also clearly pinpointed in the spectrum, as a change of behavior in the numerically-computed values (in paticular, the spectrum remains almost constant as a function of $g$ in the critical phase). In Fig.~\ref{fig:ZNTO}b we show the associated corner entropy, which clearly signals also the phase transitions. In particular, we observe that for $g>1.2$, the corner entropy depends strongly on $\chi$, which is a clear signal of the critical phase.

\begin{figure}
\includegraphics[width=0.5\textwidth]{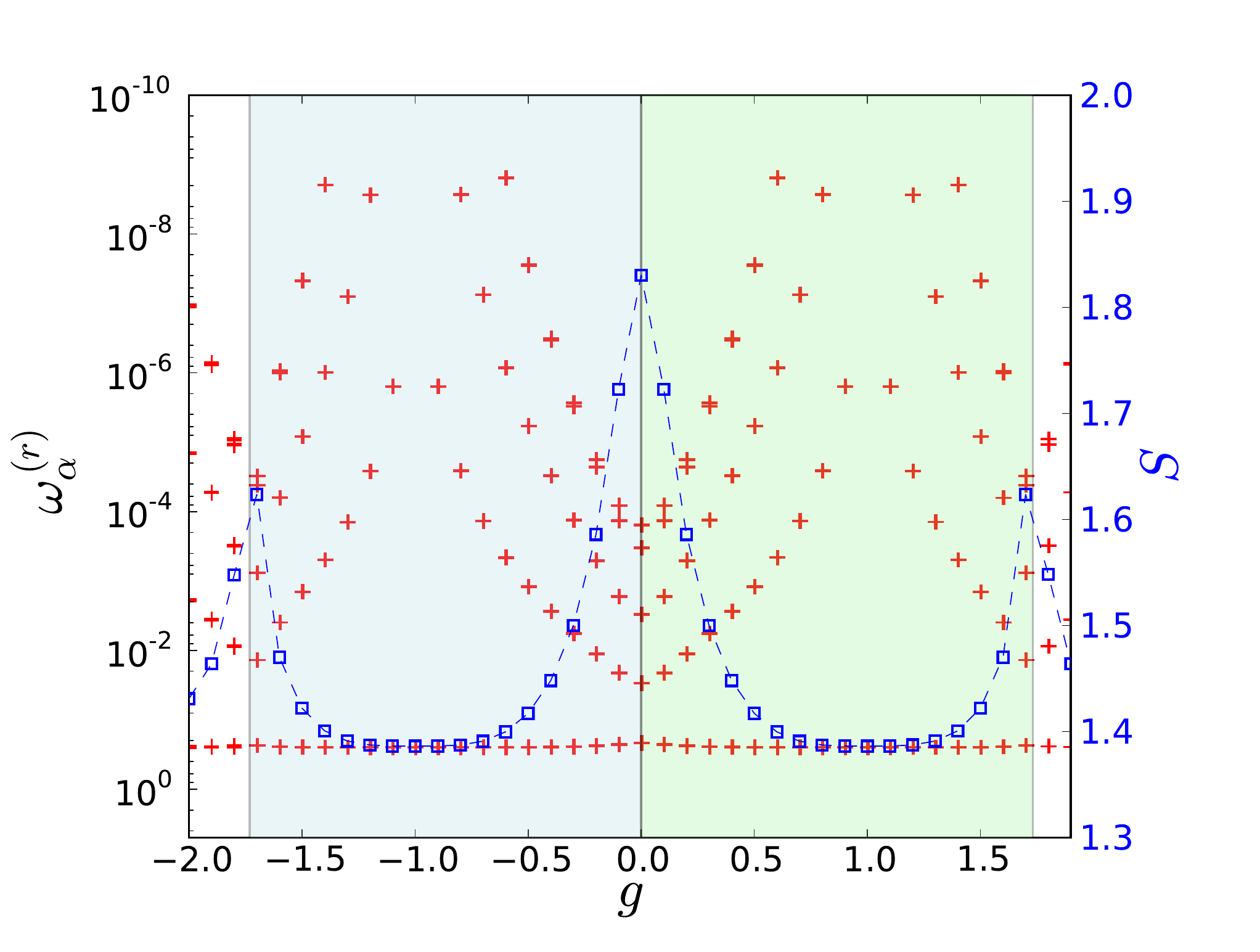}
\caption{[Color online] Corner spectra $\omega^{(r)}_\alpha$ for the norm of the $\mathbb{Z}_2$ SPT PEPS with deformation $g$ on the square lattice with $\chi=40$, together with the corner entropy computed from the corner spectra.}  
\label{fig:Z2SPT}
\end{figure}

\subsection{Perturbed SPT order}

Next, we study the quantum phase transition between two different $\mathbb{Z}_2$  symmetry-protected topological (SPT) phases on the 2d square lattice. The fixed point wave function from the 3-cocycle condition can be described by a 2d PEPS \cite{spt}, defined by a tensor $A^{i,j,k,l}_{\alpha\alpha,\beta\beta',\gamma\gamma',\delta\delta'}\equiv A[ijkl]$ satisfying $i=\alpha=\alpha'$, $j=\beta=\beta'$, $k=\gamma = \gamma'$, and $l = \delta = \delta'$, as follows:
\begin{align}
& A[0000] = A[1111] = A[0011] = A[1100] = 1  \notag \\
& A[1001] = A[0110] = A[0101] = A[1010] = 1  \notag \\
& A[0001] = A[1110] = A[0100] = A[1011] = 1  \notag \\
& A[1000] = A[0001] =  g  \notag \\
& A[0100] = A[1110] = |g|. 
\end{align}
At $g=1$, this tensor represents a fixed-point wave function for the trivial $\mathbb{Z}_2$ SPT phase. As $g=-1$, it is the fixed-point wave function of the nontrivial $\mathbb{Z}_2$ SPT phase.  
As a function of $g$, the tensor smoothly interpolates between the two phases. 
For large $|g|$ the tensor is also in an ordered phase. 

We have computed the corner spectra $\omega^{(r)}_\alpha$ and corner entropy for the doble-layer norm tensor of this state by using rCTM, which we show in Fig.~\ref{fig:Z2SPT}. We can see clearly that both the spectrum and entropy pinpoint all the phase transitions mentioned above. We find the transition to the ordered phase at $|g|=1.7$, in agreement with the results from Ref.~\cite{spt}.


\section{Chiral topological corner entanglement spectrum} 
\label{Sec7}

We have seen earlier that given a 2d Hamiltonian we can use CTs (in a 3d setup) to obtain  the entanglement spectrum of a bipartite cut separating two semi-infinite planes. We can obtain this entanglement spectrum using the 2d quantum state renormalization approach described earlier using CTs.
{ In this section,
we first consider the so-called Ising PEPS \cite{isingpeps} which, by construction, has a quantum phase transition that corresponds to the classical Ising transition, which was studied earlier in Sec.~\ref{Sec5} using the rCTM method. Here we use this state to benchmark the method, and we show the entanglement spectrum in the disordered phase. Then, we use this approach to study the boundary theory of  2d chiral topological quantum spin liquids that can be exactly described as a PEPS.

\begin{figure}
\includegraphics[width=0.475\textwidth]{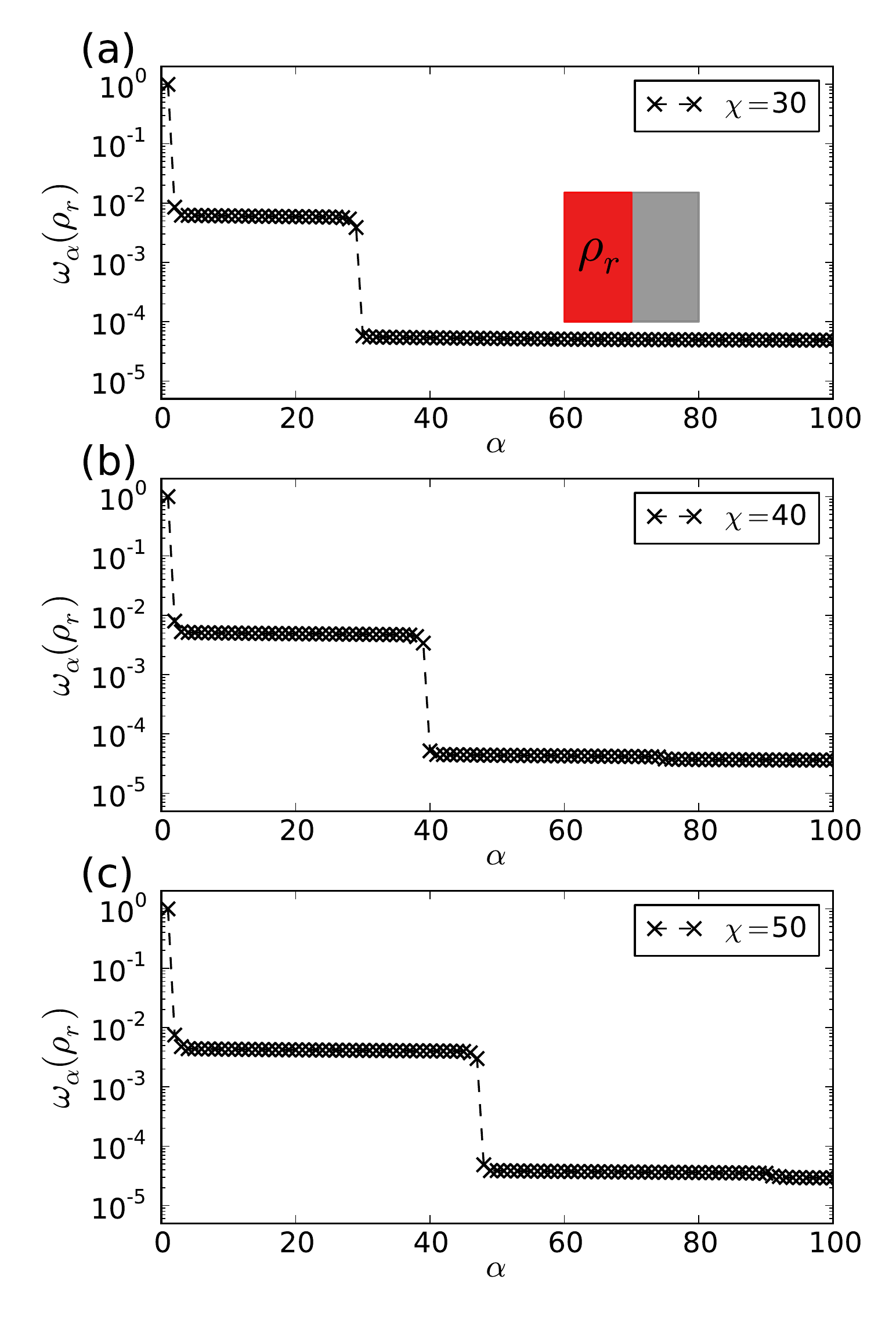}
\caption{[Color online] Entanglement pectra $\omega_{\alpha}(\rho_r)$ of  a half of 2d quantum system (see Fig.~\ref{fig:CT}) for the Ising PEPS model in disordered phase from Ref.~\cite{isingpeps}, for bond dimension (a) $\chi = 30$, (b) $\chi = 40$, and (c) $\chi = 50$.}  
\label{fig:isingPEPS}
\end{figure}

\subsection{The disorder phase: the Ising PEPS}

Let us first consider the Ising PEPS~\cite{isingpeps} on the
square lattice with tensor 
$ A=|0\rangle \langle \theta, \theta, \theta, \theta|+|1\rangle \langle \bar{\theta},\bar{\theta}, \bar{\theta}, \bar{\theta}|$, where the ket (bra) corresponds to the physical (virtual) degrees of freedom, and 
$|\theta \rangle =\cos \theta |0\rangle + \sin \theta|1\rangle$ as well as  
$|\bar{\theta} \rangle =\sin \theta |0\rangle + \cos \theta|1\rangle$ with  $\theta \in  [0,\pi/4]$. A corresponding local
Hamiltonian can be written down that has this PEPS as a ground state (not shown here)~\cite{isingpeps}.
In Ref.\cite{isingpeps} it was shown that  there is a second-order quantum phase transition from ordered phase to disorder phase occurring at $\theta_c  \approx 0.349596$. 
To illustrate that our method is not limited by the usage of corner tensors, we include results from the 2d Ising PEPS in the disorder phase with $\theta=0.5$ in Fig.~\ref{fig:isingPEPS}. 
This was studied previously in finite systems on a cylinder~\cite{isingpeps}. 
We observe that, first, there is a unique lowest entanglement eigenvalue (or one unique largest eigenvalue of corresponding transfer matrix), which is clearly identified by our method. 
Second, it is known that the low-lying entanglement spectrum seems to form one-dimensional bands (vs momentum). 
Because of the effective size introduced by the finite bond dimension, the effective momenta are discrete and we expect that our CT entanglement spectrum will see closely spaced values in one band, separated by a large gap from other bands. The number of such discrete values will depend on the bond dimension (see Fig.~\ref{fig:isingPEPS}), and the larger the bond dimension, the more points will be picked up within a band. This is exactly what we saw. }

 \begin{figure}
\includegraphics[width=0.475\textwidth]{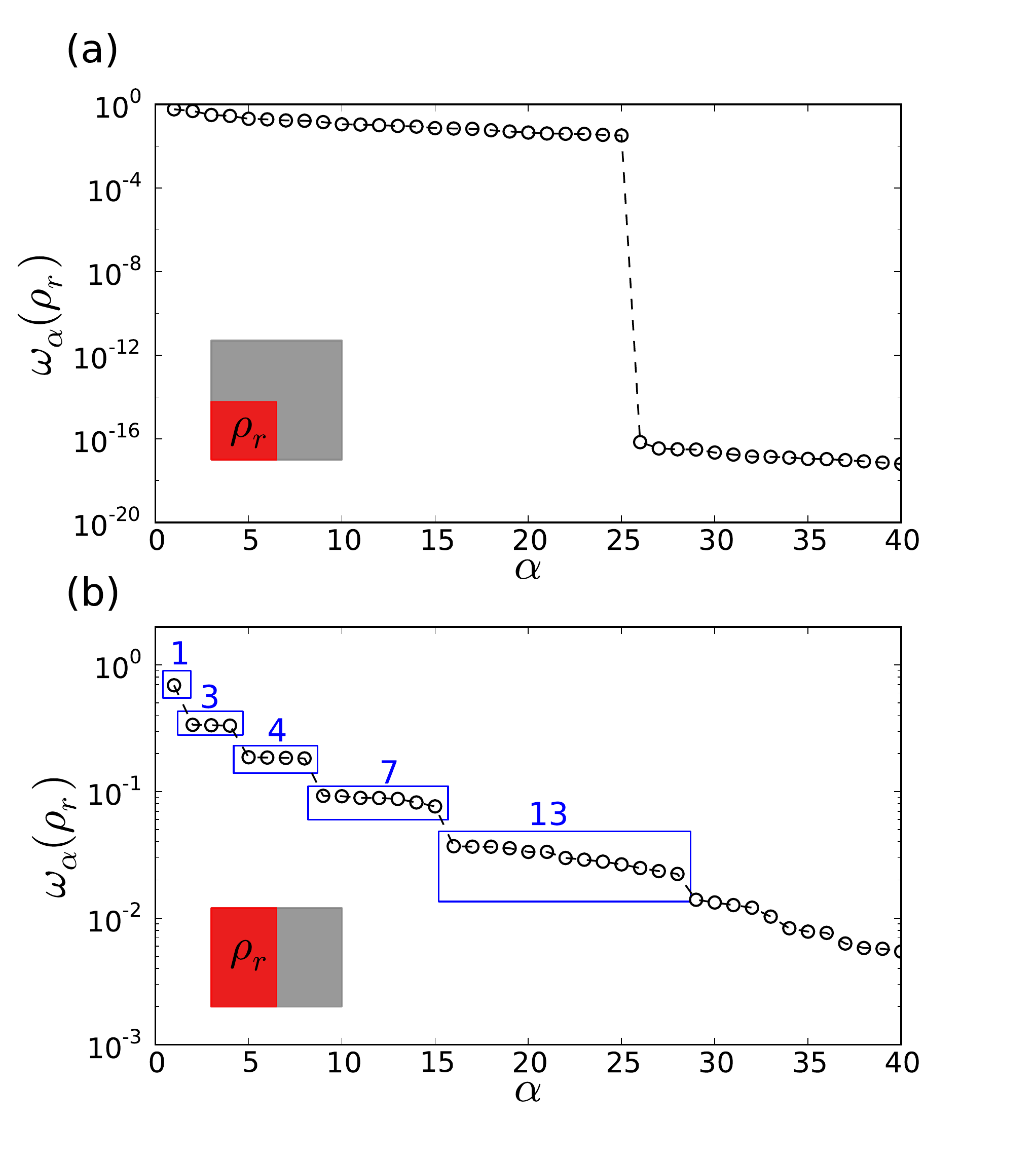}
\caption{[Color online] Entanglement pectra $\omega_{\alpha}(\rho_r)$ of (a) one quarter and (b) a half of 2d quantum system (see Fig.~\ref{fig:CT}) for the chiral topological state from Ref.~\cite{chiralPEPS}, for bond dimension $\chi = 50$. In (b) the largest spectral values are mostly converged and coincide with the expected degeneracies of the vacuum Virasoro tower of the $SU(2)_1$ WZW model describing the chiral gapless edge.}  
\label{fig:spectra-chiral}
\end{figure}

\begin{figure}
\includegraphics[width=0.475\textwidth]{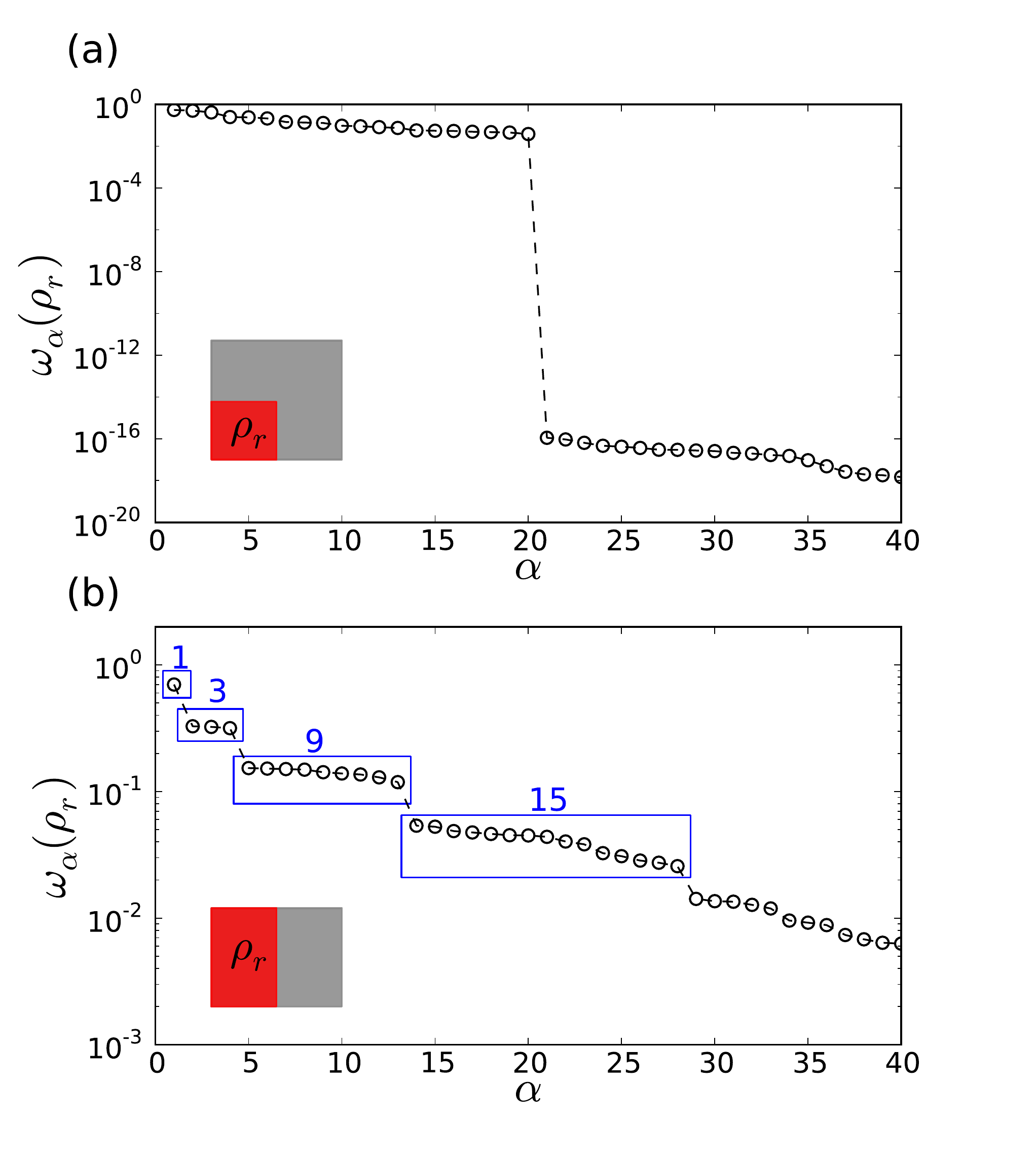}
\caption{[Color online] Entanglement pectra $\omega_{\alpha}(\rho_r)$ of (a) one quarter and (b) a half of 2d quantum system (see Fig.~\ref{fig:CT}) for the chiral topological state from Ref.~\cite{chiralPEPS}, for bond dimension $\chi = 40$. In (b) the largest spectral values are mostly converged and coincide with the expected degeneracies of the vacuum Virasoro tower of the $SU(2)_2$ WZW model describing the chiral gapless edge.}  
\label{fig:spectra-SU2_2chiral}
\end{figure}

\subsection{$SU(2)_1$ WZW chiral edge state}�
We have first studied the exact 2d PEPS with $D=3$ on a square lattice corresponding to a chiral topological quantum spin liquid with $SU(2)$ symmetry from Ref.~\cite{chiral1}.
The state is known to be critical, and has a chiral gapless edge described by a $SU(2)_1$ Wess-Zumino-Witten (WZW) CFT. The gapless edge state has been characterized previously by studying the entanglement spectrum of the PEPS on an infinitely-long but finite-circumference cylinder \cite{eHam, chiral1, chiral2}. 
In that calculation it was actually possible to find the degeneracies of the different Virasoro towers of $SU(2)_1$ corresponding to each of the highest weight states. 
If no parity or topological sector are explicitly fixed, then the numerical calculation of the entanglement spectrum naturally produces the Virasoro tower of the CFT vacuum state \cite{chiral2}.
This wave function can be given by a PEPS tensor $A^{s}_{i,j,k,l}$ with $s=\pm 1/2$ and $i,j,k,l=0,1,2$, with non-zero coefficients as follows:
\begin{align}
& A^{-1/2}_{2,0,1,1} = \!\!-\lambda_1-i \lambda_2,  
\; A^{-1/2}_{2,1,1,0} = \!\!-\lambda_1+i \lambda_2, 
\; A^{-1/2}_{2,1,0,1} = \!\!-\lambda_0 ; \notag\\
& A^{-1/2}_{1,1,2,0} = \!\! -\lambda_1-i \lambda_2,  
\; A^{-1/2}_{1,0,2,1} = \!\!-\lambda_1+i \lambda_2, 
\; A^{-1/2}_{0,1,2,1} = \!\!-\lambda_0 ; \notag\\
& A^{-1/2}_{1,2,0,1} =   \,\; \lambda_1+i \lambda_2,  
\; A^{-1/2}_{0,2,1,1} =   \,\; \lambda_1-i \lambda_2, 
\; A^{-1/2}_{1,2,1,0} =   \,\; \lambda_0 ; \notag\\
& A^{-1/2}_{0,1,1,2} =   \,\; \lambda_1+i \lambda_2,  
\; A^{-1/2}_{1,1,0,2} =   \,\; \lambda_1-i \lambda_2, 
\; A^{-1/2}_{1,0,1,2} =   \,\; \lambda_0 ; \notag\\
& A^{1/2}_{2,1,0,0} =  \,\; \lambda_1+i \lambda_2,  
\; A^{1/2}_{2,0,0,1} =   \,\; \lambda_1-i \lambda_2, 
\; A^{1/2}_{2,0,1,0} =  \,\; \lambda_0 ; \notag\\
& A^{1/2}_{0,0,2,1} =   \,\; \lambda_1+i \lambda_2,  
\; A^{1/2}_{0,1,2,0} =  \,\; \lambda_1-i \lambda_2, 
\; A^{1/2}_{1,0,2,0} =   \,\; \lambda_0 ; \notag\\
& A^{1/2}_{0,2,1,0} =  \!\!-\lambda_1-i \lambda_2,  
\; A^{1/2}_{1,2,0,0} =  \!\!-\lambda_1+i \lambda_2, 
\; A^{1/2}_{0,2,0,1} =  \!\!-\lambda_0 ; \notag\\
& A^{1/2}_{1,0,0,2} =  \!\!-\lambda_1-i \lambda_2,  
\; A^{1/2}_{0,0,1,2} =    \!\!-\lambda_1+i \lambda_2, 
\; A^{1/2}_{0,1,0,2} =  \!\!-\lambda_0,
\label{TNS_CSL}
\end{align}
where $\lambda_0=-2, \, \lambda_1=1,\text{and }   \lambda_2=1$. 

Here we have computed the entanglement spectrum of this PEPS wave function, using the quantum state renormalization approach explained previously. Our results are in Fig.~\ref{fig:spectra-chiral} for CT with a bond dimension $\chi = 50$. In the case of the entanglement spectrum for a quadrant, we see that the eigenvalues obey an almost flat distribution with a sudden drop. However, the spectrum of half an infinite system tends to obey the expected degeneracies of the Virasoro tower for the vacuum (which has angular momentum $j=0$) of the $SU(2)_1$ WZW model that describes the edge physics of this state. More specifically, the degeneracies of the 4 largest multiplets of eigenvalues are well converged and equal to $1,3,4,7$ and $13$, exactly matching the first 4 degeneracies of the Virasoro tower for the vacuum of the $SU(2)_1$ WZW model \cite{chiral1, chiral2}. We suspect the reason that we are able to see discrete spectrum rather than a continuous one is due to the effective size that the finite bond dimension introduces, even though we are using the infinite setting of the PEPS description. However, we do not see the degeneracy corresponding to the angular momentum $j=1/2$ tower.

 \subsection{$SU(2)_2$ WZW chiral edge state}

Moreover, we have considered the calculation of the entanglement spectrum from the corner properties for the double-layer chiral topological PEPS from Ref.~\cite{chiral2}, which has a gapless edge modes described by a $SU(2)_2$ WZW model. The PEPS is constructed simply from two layers of the tensors in Eq.~(\ref{TNS_CSL}) symmetrizing the physical indices (i.e., projecting in the total spin-1 subspace). Our results are in Fig.~\ref{fig:spectra-SU2_2chiral}  for CT with a bond dimension $\chi = 40$. Once again we see an almost flat spectrum with a sudden drop when we consider one quadrant. However, for half an infinite system, we see that the degeneracies of the 3 largest multiplets of eigenvalues tend to be $1,3,9$ and $15$, in agreement with the first 3 degeneracies of the Virasoro tower for the vacuum of the $SU(2)_2$ WZW model \cite{chiral2}. 

{  Furthermore, our results on chiral topological states obtained from CT agree well with the studies using cylindrical geometry~\cite{chiral1,chiral2}. In those studies as well as in ours it is found that those (discrete) degeneracy patterns show up in the low-lying entanglement spectrum and agree with the counting from conformal field theory.}

\section{Conclusions}
\label{sec:Conclusion} 

In this paper we have shown that CTMs and CTs encode universal properties of bulk physics in classical and quantum lattice systems, and that this can be computed efficiently with current state-of-the-art numerical methods. We have seen this for a wide variety of models in 1d, 2d, and 3d, both classical and quantum. First we have checked the structure of the corner energies and corner entropy for three models in the universality class of 1d quantum Ising. Then, we have used this formalism to check explicitly the correspondence between quantum systems in d dimensions and classical systems in (d+1) dimensions. In this context, we have first used the partition function approach to do this mapping, and checked numerically the correspondence for the 1d quantum Ising and quantum Potts models vs 2d classical anisotropic Ising and Potts models. Then, we have reviewed an approach by Suzuki mapping the 2d anisotropic classical Ising model to the 1d quantum XY model, and for which the corner energies and entropies showed a perfect match between the models.  For completeness we have also reviewed Peschel's approach for the quantum-classical mapping. We have also shown that corner properties can be used to pinpoint phase transitions in quantum lattice systems without the use of observable quantities. We have shown this for the 2d quantum XXZ model, perturbed 2d PEPS with $\mathbb{Z}_2$ and $\mathbb{Z}_3$ topological order, and a PEPS with perturbed SPT order. 

Perhaps more surprising is that  the corner objects can be used to obtain entanglement spectrums of 2d systems, even with chiral topological order and gapless $SU(2)_k$ edge modes, which we demonstrated for for $k=1,2$. For this we have proposed a new quantum state RG in the setting of corner matrices and tensors, which can be applied very generally to cases where the wavefunction can be written in the PEPS form. This enables efficient computation for entanglement spectrum for 2d infinite systems, which is much harder than the 1d case. Our state RG algorithm can also be straightforwardly generalized to 3d systems. All in all, we have shown that CTMs and CTs, apart from being useful numerical tools, also encode by themselves very relevant physical information that can be retrieved in a natural way from usual implementations of numerical TN algorithms. 
 
The results in this paper can be extended in a number of ways. For instance, it would be interesting to check how dynamical properties affect corner properties. A similar analysis should also be possible for dissipative systems and steady states of 2d quantum systems \cite{diss}, as well as for models with non-abelian topological order. Concerning the calculation of 2d entanglement spectra, two further considerations are in order. First, notice that one could in principle compute the ``usual" entanglement spectrum on half an infinite cylinder from the half-row and half-column tensors obtained from rCTM, wrapping them around a cylinder of finite width and proceeding as usual with the calculation of the reduced density matrix. Second, notice that a limitation of our calculation with corner tensors is that it does not provide a ``natural" way of labelling the different eigenvalues in terms of a momenta quantum number. We believe however, that this may be possible by defining appropriate translation operators on CTMs. This idea will be pursued in future works.

\acknowledgements

This work was partially supported by the National Science Foundation under Grant No. PHY 1314748 and Grant No. PHY 1620252. R.O. acknowledges the C. N. Yang Institute for Theoretical Physics for hosting him during the time that this work was initiated.


\begin{thebibliography}{99}
\bibitem{baxterCTM}
R. J. Baxter, J. Math. Phys. {\bf 9}, 650 (1968). 

\bibitem{CTMstat}
R. J. Baxter, Physica A {\bf 106}, pp18-27 (1981); R. J. Baxter, \emph{Exactly Solved Models in Statistical Mechanics} (Academic Press, London, 1982).

\bibitem{PeschelCTM}
I.  Peschel, M. Kaulke and \"O. Legeza, Ann. Physik (Leipzig) {\bf 8}, 153 (1999); I. Peschel, Braz. J. Phys. {\bf 42}, 267 (2012). 

\bibitem{dirCTM}
R. Or\'us and G. Vidal, Phys. Rev. B {\bf 80} 094403 (2009).

\bibitem{3dCT}
R. Or\'us, Phys. Rev. B {\bf 85}, 205117 (2012). 

\bibitem{BaxterVarCTM}
R.J. Baxter, J. Stat. Phys. {\bf 19} 461 (1978). 

\bibitem{KWappro}
H. A. Kramers and G. H. Wannier, Phys. Rev.  {\bf 60}, 263

\bibitem{CTMRG}
T. Nishino and K. Okunishi, J. Phys. Soc. Jpn. {\bf 65} pp. 891-894 (1996); T. Nishino and K. Okunishi, J. Phys. Soc. Jp. {\bf 66}, 3040 (1997).

\bibitem{ffUpdate}
H. N. Phien, J. A. Bengua, H. D. Tuan, P. Corboz and R. Or\'us, Phys. Rev. B {\bf 92}, 035142 (2015). 

\bibitem{frankCTM}
L. Vanderstraeten, M. Mari\"en, F. Verstraete and J. Haegeman, Phys. Rev. B {\bf 92}, 201111 (2015);  L. Vanderstraeten, J. Haegeman, P. Corboz and F. Verstraete, Phys. Rev. B {\bf 94}, 155123 (2016).

\bibitem{CT}
T. Nishino and K. Okunishi, J. Phys. Soc. Jpn. {\bf 67} 3066 (1998).

\bibitem{tn}
F. Verstraete, J. I. Cirac, and V. Murg, Adv. Phys. {\bf 57},143 (2008); 
J. I. Cirac and F. Verstraete, J. Phys. A: Math. Theor. {\bf 42}, 504004 (2009);
R. Augusiak, F. M. Cucchietti, and M. Lewenstein, in \emph{Modern Theories of Many-Particle Systems in Condensed Matter Physics}, Lect. Not. Phys. 843, 245-294 (2012); 
J. Eisert, \emph{Modeling and Simulation} {\bf 3}, 520 (2013); 
N. Schuch, QIP, Lecture Notes of the 44th IFF Spring School (2013); 
R. Or\'us, Eur. Phys. J. B {\bf 87}, 280 (2014); R. Or\'us, Ann. Phys.-New York {\bf 349} 117158 (2014).

\bibitem{eHam} 
H. Li and F. D. M. Haldane, Phys. Rev. Lett. {\bf 101}, 010504 (2008); J. I. Cirac, D. Poilblanc, N. Schuch and F. Verstraete, Phys. Rev. B {\bf 83}, 245134 (2011); S. Yang, L. Lehman, D. Poilblanc, K. Van Acoleyen, F. Verstraete, J. I. Cirac, and N. Schuch, Phys. Rev. Lett. {\bf 112}, 036402 (2014).

\bibitem{PEPS}
F. Verstraete, J. I. Cirac, cond-mat/0407066; 

\bibitem{Z2}
S. Dusuel, M. Kamfor, R. Or\'us, K. P. Schmidt and J. Vidal, Phys. Rev. Lett. {\bf 106}, 107203 (2011). 

\bibitem{Z3_1}
M. D. Schulz, S. Dusuel, R. Or\'us, J. Vidal and K. P. Schmidt, New J. Phys. {\bf 14}, 025005 (2012).

\bibitem{Z3_2}
C.-Y. Huang and T.-C. Wei, Phys. Rev. B {\bf 92}, 085405 (2015).

\bibitem{spt}
C.-Y. Huang and T.-C. Wei, Phys. Rev. B {\bf 93}, 155163 (2016).

\bibitem{chiralPEPS} 
D. Poilblanc, J. I. Cirac, and N. Schuch, Phys. Rev. B {\bf 91}, 224431 (2015). 

\bibitem{chiral1} 
D. Poilblanc, N. Schuch and I. Affleck, Phys. Rev. B {\bf 93}, 174414 (2016). 

\bibitem{chiral2}
M. Mambrini, R. Or\'us and D. Poilblanc, Phys. Rev. B {\bf 94}, 205124 (2016). 

\bibitem{IsingPEPS}
F. Verstraete, M. M. Wolf, D. P\'erez-Garc\'ia, J. I. Cirac, Phys. Rev. Lett. {\bf 96}, 220601 (2006).

\bibitem{iPEPS}
J. Jordan, R. Or\'us, G. Vidal, F. Verstraete and J. I. Cirac, Phys. Rev. Lett. {\bf 101}, 250602 (2008).

\bibitem{SuzukiXY}
M. Suzuki, Progress of Theoretical Physics, {\bf 46}, 1337 (1971). 

\bibitem{oruscorboz} 
A. Kshetrimayum, T. Picot, R. Or\'us and D. Poilblanc, Phys. Rev. B {\bf 94}, 235146 (2016); T. Picot, M. Ziegler, R. Or\'us and D. Poilblanc, Phys. Rev. B {\bf 93}, 060407 (2016); P. Corboz, Phys. Rev. B {\bf 94}, 035133 (2016); P. Corboz, Phys. Rev. B {\bf 93}, 045116 (2016); P. Corboz, T. M. Rice and M. Troyer, Phys. Rev. Lett. {\bf 113}, 046402 (2014); P. Corboz and F. Mila, Phys. Rev. Lett. {\bf 112}, 147203 (2014). 

\bibitem{ads}
K. Ueda, R. Krcmar, A. Gendiar and T. Nishino, J. Phys. Soc. Jpn. {\bf 76} 084004 (2007).

\bibitem{pbc} 
E. Bartel and A. Schadschneider, Int. J. Mod.Phys. C {\bf 19}, 81145 (2008).

\bibitem{stoch}
A. Kemper, A. Gendiar, T. Nishino, A. Schadschneider and J. Zittartz, J. Phys. A: Math. Gen. {\bf 36}, 29-41 (2003).

\bibitem{Hc}
K. Okunishi, J. Phys. Soc. Jpn. {\bf 74}, 3186-3192 (2005). 

\bibitem{Kim2016}
P. Kim, H. Katsura, N. Trivedi and J. H. Han, Phys. Rev. B {\bf 94}, 195110 (2016).

\bibitem{itebd}
G. Vidal, Phys. Rev. Lett. {\bf 98}, 070201 (2007); R. Or\'us and G. Vidal, Phys. Rev. B {\bf 78}, 155117 (2008).

\bibitem{idmrg} 
I. P. McCulloch, arXiv:0804.2509; G. M. Crosswhite, A. C. Doherty, G. Vidal, Phys. Rev. B {\bf 78}, 035116 (2008).

\bibitem{qsrg} 
F. Verstraete, J. I. Cirac, J. I. Latorre, E. Rico and M. M. Wolf, Phys. Rev. Lett. {\bf 94}, 140601 (2005). 

\bibitem{genSVD} 
Z. Y. Xie, J. Chen, M. P. Qin, J. W. Zhu, L. P. Yang and T. Xiang, Phys. Rev. B {\bf 86}, 045139 (2012). 

\bibitem{cftFreeFermion}
G. Vidal, J. I. Latorre, E. Rico and A. Kitaev, Phys. Rev. Lett. {\bf 90} 227902 (2003); J. I. Latorre, E. Rico and G. Vidal, Quant. Inf. and Comp. {\bf 4}, 48-92 (2004); V. Korepin, Phys. Rev. Lett. {\bf 92} 096402 (2004); A. R. Its, B. Q. Jin, and V. E. Korepin, Journal Phys. A:Math. Gen. {\bf 38}, 2975-2990, (2005); P. Calabrese and J. Cardy, JSTAT 0406:002 (2004).

\bibitem{mps}
M. Fannes, B. Nachtergaele and R. F. Werner, Commun. Math. Phys. {\bf 144}, 443-490 (1992); A. Kl\"umper, A. Schadschneider, J. Zittartz, J. Phys. A {\bf 24}, L955 (1991); A. Kl\"umper, A. Schadschneider, J. Zittartz, Europhys. Lett. {\bf} 24, 293 (1993).

\bibitem{partition}
See, e.g., S. Sachdev, \emph{Quantum Phase Transitions}, Cambridge University Press, 2nd Edition (2011).

\bibitem{PeschelMap}
I. Peschel, Physics Letters A, {\bf 110}, 313 (1985).

\bibitem{BloteDeng}
H. W. J. Bl\"ote and Y. Deng
Phys. Rev. E {\bf 66}, 066110 (2002).
\bibitem{XYmodel}
M. Henkel, Conformal Invariance and Critical Phenomena (Springer, Berlin, 1999).

\bibitem{2dXYmodel}
M. Henkel, Journal of Physics A: Mathematical and General, {\bf 17}, L795 (1984)

\bibitem{BarouchMcCoy}
E. Barouch and B. M. McCoy, Phys. Rev. A {\bf 3}, 786 (1971).

\bibitem{fidelity}
P. Zanardi and N. Paunkovic, Phys. Rev. E {\bf 74}, 031123 (2006); 
H.-Qiang Zhou and J.P. Barjaktarevic, cond-mat/0701608; H.-Qiang Zhou, J.-Hui Zhao and B. Li, arXiv:0704.2940;
H.-Qiang Zhou, arXiv:0704.2945; P. Zanardi, M. Cozzini and P. Giorda, cond-mat/0606130; N. Oelkers and J. Links, Phys. Rev. B {\bf 75}, 115119 (2007); M. Cozzini, R. Ionicioiu and P. Zanardi, cond-mat/0611727; L. Campos Venuti and P. Zanardi, Phys. Rev. Lett. {\bf 99}, 095701 (2007); P. Buonsante and A. Vezzani, Phys. Rev. Lett. {\bf 98}, 110601 (2007); W.-L. You, Y.-W. Li and S.- J. Gu, Phys. Rev. E {\bf 76}, 022101 (2007); S.-J. Gu et al., Phys. Rev. B {\bf 77}, 245109 (2008); M.-F. Yang, Phys. Rev. B {\bf 76}, 180403(R) (2007); Y.-C. Tzeng and M.-F. Yang, Phys. Rev. A {\bf 77}, 012311 (2008); H.-Qiang Zhou, R. Or\'us and G. Vidal, Phys.Rev. Lett. {\bf 100}, 080601 (2008); P. Schmoll and R. Or\'us, arXiv:1605.04315. 

\bibitem{xxz}
M. Kohno and M. Takahashi, Phys. Rev. B {\bf 56}, 3212 (1997); S. Yunoki, Phys. Rev. B {\bf 65}, 092402 (2002).

\bibitem{su}
H. C. Jiang, Z. Y. Weng, T. Xiang, Phys. Rev. Lett. {\bf 101}, 090603 (2008).

\bibitem{TC}
A. Y. Kitaev, Ann. Phys. 303, {\bf 2} (2003).

\bibitem{Z2_deform}
X. Chen, Z.-C. Gu and X.-G. Wen, Phys. Rev. B {\bf 82}, 155138 (2010).

\bibitem{RK}
D. S. Rokhsar and S. A. Kivelson, Phys. Rev. Lett. {\bf 61}, 2376 (1988).

\bibitem{diss}
A. Kshetrimayum, H. Weimer and R. Or\'us, arXiv:1612.00656. 


\bibitem{isingpeps}
M. Rispler, K. Duivenvoorden, and N. Schuch, Phys. Rev. B {\bf 92}, 155133 (2015); 
F. Verstraete, M. M. Wolf, D. Perez-Garcia, and J. I. Cirac, Phys. Rev. Lett. {\bf 96}, 220601 (2006).

\bibitem{ctmnote}
http://quattro.phys.sci.kobe-u.ac.jp/nishi/Note/Beijing1.pdf


\end{thebibliography}
 \end{document}